\documentclass[preprint,authoryear,12pt]{elsarticle}

\usepackage[german,american]{babel}

\usepackage{times}
\usepackage{helvet}
\usepackage{latexsym}

\usepackage{bm}
\usepackage{amsfonts}
\usepackage{amssymb}
\usepackage{amsmath}
\usepackage{booktabs}

\usepackage{isomath}
\usepackage[title]{appendix}
\usepackage{miller}

\usepackage[usenames]{color}
\definecolor{darkgreen}{rgb}{0.0,0.5,0.0}

\usepackage{siunitx}
\usepackage{tikz} 
\usepackage{pgfplots}
\usepackage{graphicx}
\usepackage{subfigure}
\usepackage[percent]{overpic}
\usepgfplotslibrary{units}

\pgfplotsset{compat=newest} 
\usepgfplotslibrary{units} 
\graphicspath{{./figures/}}

\usepackage{float}

\usepackage{caption}

\usepackage[noabbrev]{cleveref}
\usepackage[linesnumbered, ruled]{algorithm2e}

\usepackage{epic,eepic}

\newcommand{\ie}{\textit{i.e.}}

\newcommand{\Euler}{\text{Euler}}

\newcommand{\PK}{\text{Piola-Kirchhoff}}
\newcommand{\Mandel}{\text{Mandel}}

\newcommand{\identity}{\ensuremath{\tnsr I}}

\newcommand{\transpose}[1]{\ensuremath{{#1}^{\text T}}}
\newcommand{\inverse}[1]{\ensuremath{{#1}^{-1}}}
\newcommand{\invtranspose}[1]{\ensuremath{{#1}^{\text{-T}}}}
\newcommand{\sign}[1]{\ensuremath{\operatorname{sgn}\left({#1}\right)}}

\newcommand{\Grad}[1][]{\ensuremath{\nabla{#1}}}

\newcommand{\Divergence}[1][]{\ensuremath{\operatorname{Div}{#1}}}

\newcommand{\inc}[1]{\ensuremath{\text d{#1}}}
\newcommand{\abs}[1]{\ensuremath{\left|{#1}\right|}}
\newcommand{\norm}[2][]{\ensuremath{\left|\left|{#2}\right|\right|\if\relax\detokenize{#1}\relax\else _{#1}\fi}}

\newcommand{\domain}[1]{\ensuremath{\mathcal{#1}}}
\newcommand{\tnsrfour}[1]{\ensuremath{\mathbb{#1}}}
\newcommand{\tnsr}[1]{\ensuremath{\mathbf{#1}}}

\newcommand{\vctr}[1]{\ensuremath{\mathbf{#1}}}
\newcommand{\vctrgreek}[1]{\ensuremath{\bm{#1}}}

\newcommand{\eyetwo}{\ensuremath{\tnsr I}}

\newcommand{\F}{\ensuremath{\tnsr F}}
\newcommand{\Fdot}{\ensuremath{\dot{\tnsr F}}}
\newcommand{\Fp}[1][]{\ensuremath{\tnsr F_\text{p#1}}}

\newcommand{\Fpdot}[1][]{\ensuremath{\dot{\tnsr F}_\text{p#1}}}
\newcommand{\Fc}[1][]{\ensuremath{\tnsr F_\text{c#1}}}

\newcommand{\Fe}{\ensuremath{\tnsr F_\text{e}}}

\newcommand{\Lp}[1][]{\ensuremath{\tnsr L_\text{p#1}}}

\newcommand{\GL}{\ensuremath{\tnsr E_\text{e}}}

\newcommand{\fPK}{\ensuremath{\tnsr P}}
\newcommand{\sPK}{\ensuremath{\tnsr S}}

\newcommand{\Mp}[1][]{\ensuremath{\tnsr M_\text{p#1}}}

\newcommand{\defmap}{\ensuremath{\vctrgreek{\chi}}}

\newcommand{\ese}[1][]{\mathcal{W}_e}
\newcommand{\esec}[1][]{{\mathcal{W}}^{crit}_e}

\newcommand{\X}[1][]{\ensuremath{\vctr X}}
\newcommand{\x}[1][]{\ensuremath{\vctr x}}
\newcommand{\disp}[1][]{\ensuremath{\vctr u}}

\newcommand{\dX}[1][]{\ensuremath{\vctr{dX}}}
\newcommand{\dx}[1][]{\ensuremath{\vctr{dx}}}
\newcommand{\dy}[1][]{\ensuremath{\vctr{dy}}}
\newcommand{\du}[1][]{\ensuremath{\vctr{du}}}
\newcommand{\dv}[1][]{\ensuremath{\vctr{dv}}}

\newcommand{\vecx}[1][]{\ensuremath{\vctr{x}}}
\newcommand{\vecX}[1][]{\ensuremath{\vctr{X}}}
\newcommand{\vecy}[1][]{\ensuremath{\vctr{y}}}
\newcommand{\vecu}[1][]{\ensuremath{\vctr{u}}}
\newcommand{\vecv}[1][]{\ensuremath{\vctr{v}}}

\newcommand{\invF}[1][]{\ensuremath{\inverse{\F}}}
\newcommand{\tranF}[1][]{\ensuremath{\transpose{\F}}}
\newcommand{\invtranF}[1][]{\ensuremath{\invtranspose{\F}}}

\newcommand{\pl}{\textsubscript{p}}
\newcommand{\da}{\textsubscript{d}}
\newcommand{\el}{\textsubscript{e}}

\newcommand{\al}{\textsuperscript{\alpha}}

\newcommand{\vecnda}[1][]{\ensuremath{\hat{\vctr{n}}\da{}\al}}

\newcommand{\vecn}[1][]{\ensuremath{\hat{\vctr{n}}}}
\newcommand{\vecd}[1][]{\ensuremath{\hat{\vctr{d}}}}
\newcommand{\vect}[1][]{\ensuremath{\hat{\vctr{t}}}}

\newcommand{\vecnp}[1][]{\ensuremath{\hat{\vctr{n}}\pl}}
\newcommand{\vecdp}[1][]{\ensuremath{\hat{\vctr{d}}\pl}}
\newcommand{\vectp}[1][]{\ensuremath{\hat{\vctr{t}}\pl}}

\newcommand{\vecnd}[1][]{\ensuremath{\hat{\vctr{n}}\da}}
\newcommand{\vecdd}[1][]{\ensuremath{\hat{\vctr{d}}\da}}
\newcommand{\vectd}[1][]{\ensuremath{\hat{\vctr{t}}\da}}

\newcommand{\vecne}[1][]{\ensuremath{\hat{\vctr{n}}\el}}
\newcommand{\vecde}[1][]{\ensuremath{\hat{\vctr{d}}\el}}
\newcommand{\vecte}[1][]{\ensuremath{\hat{\vctr{t}}\el}}

\newcommand{\invFe}[1][]{\ensuremath{\inverse{\Fe}}}
\newcommand{\tranFe}[1][]{\ensuremath{\transpose{\Fe}}}
\newcommand{\invtranFe}[1][]{\ensuremath{\invtranspose{\Fe}}}

\newcommand{\invFp}[1][]{\ensuremath{\inverse{\Fp}}}
\newcommand{\tranFp}[1][]{\ensuremath{\transpose{\Fp}}}
\newcommand{\invtranFp}[1][]{\ensuremath{\invtranspose{\Fp}}}

\newcommand{\invFd}[1][]{\ensuremath{\inverse{\Fc}}}
\newcommand{\tranFd}[1][]{\ensuremath{\transpose{\Fc}}}
\newcommand{\invtranFd}[1][]{\ensuremath{\invtranspose{\Fc}}}

\newcommand{\svsc}[1][]{\ensuremath{\xi}}
\newcommand{\svdsc}[1][]{\ensuremath{\xi\da}}
\newcommand{\svpsc}[1][]{\ensuremath{\xi\pl}}

\newcommand{\sv}[1][]{\ensuremath{\vctrgreek \xi}}
\newcommand{\svd}[1][]{\ensuremath{\sv\da}}
\newcommand{\svp}[1][]{\ensuremath{\sv\pl}}

\newcommand{\dotsv}[1][]{\ensuremath{\dot{\sv}}}
\newcommand{\dotsvd}[1][]{\ensuremath{\dot{\svd}}}
\newcommand{\dotsvp}[1][]{\ensuremath{\dot{\svp}}}

\newcommand{\deltasv}[1][]{\ensuremath{\Delta\sv}}
\newcommand{\deltasvd}[1][]{\ensuremath{\Delta\svd}}
\newcommand{\deltasvp}[1][]{\ensuremath{\Delta\svp}}

\newcommand{\y}[1][]{\ensuremath{\vctr y}}

\newcommand{\ShearDir}[1]{\ensuremath{\vctr s^{#1}}}
\newcommand{\n}[1]{\ensuremath{\vctr n^{#1}}}

\newcommand{\dotgalpha}{\ensuremath{\dot{\gamma}^{\alpha}}}

\journal{}

\begin{document} 

\begin{frontmatter}

\title{Multi-component chemo-mechanics based on transport relations for the chemical potential}

\author[UM,MPIE]{P.~Shanthraj\corref{PS}}
\ead{pratheek.shanthraj@manchester.ac.uk}
\author[MPIE]{C.~Liu\corref{PS}}
\ead{c.liu@mpie.de}
\author[MPIE,MM]{A.~Akbarian}
\author[MPIE,MM]{B.~Svendsen}
\author[MPIE]{D.~Raabe}

\cortext[PS]{Corresponding author}
\address[UM]{The School of Materials, The University of Manchester, Manchester M1 3BB, UK}
\address[MPIE]{Max-Planck-Institut f\"ur Eisenforschung, Max-Planck-Stra{\ss}e 1, 40237 D\"usseldorf, Germany}
\address[MM]{Material Mechanics, RWTH Aachen University, Schinkelstra{\ss}e 2, 52062 Aachen, Germany}
\journal{Computer Methods in Applied Mechanics and Engineering}
\begin{abstract}

A chemo-mechanical model for a finite-strain elasto-viscoplastic material containing multiple chemical components is formulated and an efficient numerical implementation is developed to solve the resulting transport relations.
The numerical solution relies on inverting the constitutive model for the chemical potential.
In this work, a semi-analytical inversion for a general family of multi-component regular-solution chemical free energy models is derived.
This is based on splitting the chemical free energy into a convex contribution, treated implicitly, and a non-convex contribution, treated explicitly.
This results in a reformulation of the system transport equations in terms of the chemical potential rather than the composition as the independent field variable.
The numerical conditioning of the reformulated system, discretised by finite elements, is shown to be significantly improved, and convergence to the Cahn-Hilliard solution is demonstrated for the case of binary spinodal decomposition.
Chemo-mechanically coupled binary and ternary spinodal decomposition systems are then investigated to illustrate the effect of anisotropic elastic deformation and plastic relaxation of the resulting spinodal morphologies in more complex material systems.

\end{abstract}


\end{frontmatter}

\section{Introduction}
\label{sec:Int}

The main goal of modern materials science is the theory-guided tailoring of materials, including chemical composition and microstructure, in order to obtain improved properties for a sustainable technological development. 
While there has been a tremendous growth over recent years in the use of modelling and simulation tools towards this goal \citep{Shanthraj2012,Wu2014,Liu2018,Roters2019}, the realistic prediction of the thermo-chemo-mechanical interactions relevant to industrial processes is still a key development required to enable technological advances in material design, manufacturing and product development for harsh-service environments.
Among several full-field simulation approaches, the phase-field method is particularly well-suited to model interface kinetics \citep{Boettinger2002,Chen2002,chen2002phase,Moelans2008,Emmerich2008,steinbach2009phase}.
It has been successfully applied to describe many thermo-chemo-mechanical processes including solidification \citep{Nestler2000}, precipitation \citep{Zhou2010,Jokisaari2017,Rudraraju2016}, fracture \citep{spatschek2011phase,Schneider2016,Shanthraj2016,Shanthraj2017} and dislocation motion \citep{wang2001nanoscale,shen2003phase,Mianroodi2019}.
Further methodological developments including chemo-mechanical interface modelling and homogenisation have been treated recently in \cite{Svendsen2018}.

The use of diffuse-interface models to describe interfacial phenomena dates back to \citet{Cahn1958}.
The original Cahn-Hilliard (CH) equation was used to model spinodal decomposition in binary alloys, but has since been extended to multi-component systems and coupled with microelasticity \citep{hu2001phase,fischer2003kinetics,garcke2005cahn,steinbach2006multi,Moelans2008}.
A critical challenge in simulating the thermodynamics of multi-component chemo-mechanical systems is the numerical approximation of a generally non-convex chemical free energy.
Numerical methods have been developed to solve such systems using finite difference \citep{Furihata2001}, mixed finite element with composition and chemical potential treated as independent fields \citep{Barret1999,Gomez2011}, isogeometric \citep{Gomez2008} and spectral \citep{Zhu1999} spatial discretisations.
In the context of numerical time-integration, the stability, robustness and efficiency of the resulting solution  algorithm is sensitive to the non-convexity of the chemical free energy. 
A successful approach in this regard is the splitting of the chemical free energy into concave and convex components, which are then approximated separately \citep{Elliot1993,Eyre1998}. 
In particular, splitting methods have recently been applied to the multi-component, multi-phase CH equation with a generalized non-convex Landau energy \citep{Boyer2011,Tavakoli2016}. 
In the context of a mixed finite-element-based weak formulation of the classic CH relation, \citet{Gomez2011} employed a chemical energy splitting approach based on the sign of the fourth-order derivatives of the underlying energy functional.
A recent mixed weak formulation of chemo-mechanics for two-phase, two-component finite-deformation gradient elastic solids based on unconditionally stable, second-order accurate time-integration and Taylor expansion of the non-convex energy has been given by \cite{Sagiyama2016}. 

More recently, starting from a grand potential functional \citep{Plapp2011,Choudhury2012}, the transport relations have been formulated for multi-component and multi-phase systems \citep{Aagesen2018}.
In such an approach, the thermodynamics of the system is set in a grand canonical ensemble and formulated in terms of the chemical potential.
As a result, boundary and interface conditions based on the chemical potential \citep{Kim1999}, needed for a realistic representation of macroscopic systems where the total number of particles scannot be fixed, are naturally incorporated.
However, these approaches are based on a Legendre transformation, which does not exist for more general non-convex forms of the chemical free energy. 
In this work, a convex splitting of the chemical free energy is instead used to invert the multi-component chemical potential\textendash composition relation.
This results in an expression for component transport in terms of the chemical potential rather than the chemical composition, analogous to the grand potential approach, but applicable to more general forms of the chemical free energy. 
This paper is organised as follows: the basic model formulation for a single phase elasto-viscoplastic solid is presented in Section 2, followed by an outline of its numerical implementation in Section 3. 
In particular, since a dissipation potential exists for the flux-force relations of the current model, the corresponding initial boundary-value problems (IBVP) can be formulated with the help of rate variational methods \citep{Sve04,Miehe2011,Miehe2014}.
In Section 4 representative examples are used to benchmark and compare the proposed method with conventional transport relations. 
A summary is provided in Section 5 along with perspectives for future applications.

\section{Model formulation}

\subsection{Basic relations}

Let $\domain B_0 \subset \tnsrfour R^3$ be a microstructural domain of interest, with boundary $\partial \domain B_0$.
Attention is restricted in this work to the case of a single-phase elasto-viscoplastic solid mixture of $M$ diffusing chemical components.
Component diffusion and reaction in the mixture is modeled as usual by the corresponding component mass or number balance relations 
\begin{equation}
\dot{c}_m = - \Divergence \vctr{j}_m + \sigma_m, \qquad \text{for} \quad m = 1, \ldots M,
\label{equ:RelTraCom}
\end{equation} 
in the mixture in terms of component (mass or number) concentration $0 \le c_m \le 1$, the corresponding flux 
density $\vctr{j}_m$, and the corresponding specific supply-rate, $\sigma_m$. 
Assuming the mixture is closed with respect to constituent mass or number, the corresponding sum relations 
\begin{equation}
1 =\sum\limits_{m=1}^{M}c_{m}
\,,\quad
\vctr{0} =\sum\limits_{m=1}^{M}\vctr{j}_m
\,,\quad 
0 =\sum\limits_{m=1}^{M}\sigma_m,
\label{equ:RelSumCom}
\end{equation} 
and \eqref{equ:RelTraCom} imply $\sum_{m=1}^{M}\dot{c}_{m}=0$.

Besides this, the deformation resulting from an applied loading defines a field $\defmap(\vctr x):\vctr x \in \domain B_0 \to \vctr y \in \domain B$ mapping points \vctr x in the undeformed configuration $\domain B_0$ to points \vctr y in the deformed configuration \domain B.
Following \citet{Shanthraj2017}, the total deformation gradient, given by $\F = \partial\defmap/\partial\vctr x = \Grad \defmap$, is multiplicatively decomposed as 
\begin{equation}
\label{eq:muldecomp}
\F = \Fe\,\Fc\,\Fp,
\end{equation}
where \Fp\ is a lattice preserving isochoric mapping due to plastic deformation, \Fc\ represents the local deformation due to solute misfit, and \Fe\ is a mapping to the deformed configuration.
In the current approach the stress relaxation due to the component diffusion process is captured through the stress-free deformation gradient, \Fc.

Restricting attention to isothermal and quasi-static processes with no external supplies of momentum or energy, the balance relations for linear momentum, angular momentum, and internal energy, are given by
\begin{equation}
\bm{0} = \Divergence\,\fPK
\,,\quad
\transpose{\fPK}\F =  \fPK\transpose{\F}
\,,\quad
\dot{\varepsilon}
=\fPK\cdot\Fdot,
\label{equ:BalMomEneEntDam}
\end{equation}
where, \fPK\ is the first \PK\ stress, and $\varepsilon$ is the referential internal energy density.
As well the balance relation, 
\begin{equation}
\dot{\eta} =\theta^{-1}\delta- \theta^{-1} \Divergence\sum\limits_{m=1}^{M}\mu_{m}\vctr{j}_{m}\, + \theta^{-1}\sum\limits_{m=1}^{M}\mu_{m}\sigma_{m},
\label{equ:BalEneMix}
\end{equation}
for the entropy holds \cite[e.g.,][Chapter 2]{deG62}, 
with dissipation-rate density, $\delta$, absolute temperature, $\theta$, and the chemical potential, $\mu_{m}$, of component $m$. 

Combination of \eqref{equ:RelTraCom}-\eqref{equ:BalEneMix} yields the form 
\begin{equation}
\delta =\fPK\cdot\Fdot + \sum\limits_{m=1}^{M-1} \tilde{\mu}_{m}\dot{c}_{m} - \dot{\psi} - \sum\limits_{m=1}^{M-1}\vctr{j}_{m}\cdot\nabla\tilde{\mu}_{m}
\label{equ:DenRatDisMix}
\end{equation} 
for the dissipation-rate density in terms of $\tilde{\mu}_{a}:=\mu_{a}-\mu_{M}$, and the free energy density of the mixture, $\psi:=\varepsilon-\theta\eta$.

The current model formulation is based on the basic constitutive form 
\begin{equation}
\psi\left(\F, \Fp, \Fc, c_m, \tilde{c}_m, \Grad \tilde{c}_m\right) = \psi_{\mathrm{e}}\left(\F, \Fp, \Fc\right) + \psi_{\mathrm{c}}\left(c_m, \tilde{c}_m, \Grad \tilde{c}_m\right)\,,
\label{equ:DenEneFreGen}
\end{equation}
for $\psi$  in terms of elastic, $\psi_\mathrm{e}$, and chemical, $\psi_\mathrm{c}$, parts.
A non-local field, $\tilde{c}_m$ for independent components, $m = 1, \ldots M-1$, is introduced to weakly enforce any dependence of the energy on chemical inhomogeneity through its gradient $\Grad \tilde{c}_m$ \citep{Ubachs2004}. 
Modelling \fPK\ and $\tilde{\mu}_{m}$ as purely energetic, \ie\ work conjugate to \F\ and $c_m$ respectively, results in the constitutive relations
\begin{equation}
\tilde{\mu}_{m} = \partial_{c_m}\psi + \partial_{\Fc}\psi \cdot \partial_{c_m} \Fc \,,\quad\fPK = \partial_{\F }\psi.
\label{equ:StsPKF}
\end{equation}
Together, \eqref{equ:DenRatDisMix}-\eqref{equ:StsPKF} result in the residual form 
\begin{equation}
\delta = - \partial_{\Fp} \psi \cdot \Fpdot - \sum\limits_{m=1}^{M-1}\vctr{j}_{m}\cdot\nabla\tilde{\mu}_{m}
\label{equ:DenRatDisRes}
\end{equation}
for the dissipation-rate density.

\subsection{Energetic constitutive relations} 
\label{sec: constitutive}

The elastic energy, $\psi_{\mathrm{e}}$, is modelled here relative to the intermediate configuration by the form 
\begin{equation}
\psi_{\mathrm{e}}(\F,\Fp,\Fc) = \frac{1}{2}\,\GL \cdot \tnsrfour{C}\,\GL,
\label{equ:DenEneFreElaDamUn}
\end{equation}
in terms of the Green-Lagrange strain measure 
\begin{equation}
\GL = \frac{1}{2}\transpose{\Fc}\left(\transpose{\Fe}\Fe - \identity \right)\Fc,
\end{equation} 
and anisotropic elastic stiffness $\tnsrfour{C}$. 
The work conjugate second \PK\ stress is given by,
\begin{equation}
\label{eq:GLstrainP}
\sPK = \tnsrfour{C} \GL.
\end{equation}
The 1st \PK\ stress tensor, \fPK, is then related to \sPK\ through 
\begin{align}
\label{eq:2pkto1pk}
\fPK = \Fe\,\Fc\,\sPK\,\invtranspose{\Fp}.
\end{align}
Further details are provided in \citet{Roters2019}.

\subsubsection{Crystal plasticity}
\label{sec: Phenomenological plasticity}

The plastic deformation gradient is given in terms of the plastic velocity gradient, \Lp, by the flow rule
 \begin{equation}
\label{eq:plastic flow rule}
\Fpdot = \Lp \Fp,
\end{equation}
where \Lp\ is work conjugate with the \Mandel\ stress in the plastic configuration, 
\begin{equation}
\label{eq:mandelP}
\Mp = - \partial_{\Fp} \psi \transpose{\Fp} = \transpose{(\Fe\Fc)} \Fe\Fc \sPK \approx \transpose{\Fc} \Fc \sPK,
\end{equation}
assuming small elastic strains. 
A crystal plasticity model is used, where the plastic velocity gradient \Lp\ is composed of the slip rates $\dot\gamma^\alpha$ on crystallographic slip systems, which are indexed by $\alpha$ 
\begin{equation}
\label{eqn:plasticFlowRule}
\Lp = \Fpdot\inverse{\Fp} = \sum_{\alpha} \dotgalpha \,  \ShearDir{\alpha}\otimes\n{\alpha},
\end{equation}
where \ShearDir{\alpha} and \n{\alpha} are unit vectors along the slip direction and slip plane normal, respectively \citep{Roters2010,Shanthraj2011}.
The slip rates are given by the phenomenological description of \citet{Pei1983},
\begin{equation}
\label{equ:RulFloSysGli}
\dot\gamma^\alpha = \dot\gamma_0\abs{\frac{\tau^\alpha}{g^\alpha}}^n \sign{\tau^\alpha},
\end{equation}
in terms of the reference shear rate $\dot\gamma_0$, and strain-rate sensitivity exponent $n$.
The slip resistances on each slip system, $g^\alpha$, evolve asymptotically towards $g_\infty$ with shear $\gamma^\beta$ ($\beta = 1,\ldots,12$) according to the relationship
\begin{align}
\label{eq: hardening pheno}
\dot g^\alpha &=  \dot\gamma^\beta \, h_0 \left| 1 - g^\beta/g_\infty \right|^a \sign{1 - g^\beta/g_\infty}  h_{\alpha\beta},
\end{align}
with parameters $h_0$ and $a$.
The interaction between different slip systems is captured by the hardening matrix $h_{\alpha\beta}$.

The plastic dissipation can then be reduced to the simpler slip system based conjugate pair
\begin{equation}
\label{equ:plastdis}
 - \partial_{\Fp} \psi \cdot \Fpdot =  \sum_{\alpha} \tau^\alpha\ \dot\gamma^\alpha, \quad\text{where}\quad  \tau^\alpha = \Mp \cdot (\ShearDir{\alpha}\otimes\n{\alpha}).
\end{equation}
On this basis, non-negativity of $\dot\gamma_0$, $g_\infty$, $h_0$ and $h_{\alpha\beta}$ is sufficient to ensure non-negative plastic dissipation. 

\subsubsection{Multi-component chemo-mechanics}

Following \citet{Huter2018}, the local deformation arising from volumetric mismatch between solute components is given by
 \begin{equation}
\label{eq:inelastic flow rule}
\Fc = \sum\limits_{m=1}^{M-1}\left(1+\nu_{m}{c}_{m}\right)\eyetwo,
\end{equation}
where the volumetric mismatch coefficient, $\nu_{m}$, is related to the Vegard's coefficient, \ie\ $\nu_{m} = V_m/a_M$, where $V_m$ is the Vegard's coefficient for component $m$ dissolved as a solid solution in a matrix of component $M$, and  $a_M$ is the lattice parameter of the matrix.

The chemical energy, $\psi_{\mathrm{c}}$, is modelled here using the following regular solution form 
\begin{align}
\Omega \psi_{\mathrm{c}}(c_m, \tilde{c}_m, \Grad \tilde{c}_m) = &\sum_{m=1}^{M-1} E_{m}^{sol}c_m + \sum_{m,n=1}^{M-1} E_{mn}^{int}c_m c_n + R \theta \sum_{m=1}^{M} c_m\ln c_m \nonumber\\
 & + \frac{1}{2}\sum_{m=1}^{M-1} \alpha_m \left(c_m - \tilde{c}_m\right)^2+ \frac{1}{2}\sum_{m=1}^{M-1}\kappa_m |\Grad \tilde{c}_m|^2,
\label{equ:DenEneFreCheDamUn}
\end{align}
where $\Omega$ is the molar volume; $E_{m}^{sol}$, $\kappa_m$  and $\alpha_m$ are respectively the solution energy, gradient coefficient, and penalty parameter to weakly enforce $c_m = \tilde{c}_m$, for component $m$; $E_{mn}^{int}$ is the interaction energy between component $m$ and $n$; and $R$ is the universal gas constant.
The non-local field, $\tilde{c}_m$, is obtained through the equilibrium relation,
\begin{align}
0 &= \Omega \partial_{\tilde{c}_m} \psi = \Omega \left(\partial_{\tilde{c}_m} \psi 
- \Divergence \partial_{\nabla\tilde{c}_m}\psi\right) \nonumber\\
&= \alpha_m \left(\tilde{c}_m - c_m\right) - \Divergence \kappa_m \Grad \tilde{c}_m + \alpha_m \left(c_m - \tilde{c}_m\right).
\label{equ:NonLocalEq}
\end{align}

Following \citet{Onsager1931}, a linear flux-force form is assumed for the component diffusion,
\begin{equation}
\vctr{j}_{m} = -M_m\Grad\tilde{\mu}_{m},
\label{equ:FluCom}
\end{equation}
with $M_m$ the component mobility in the lattice-fixed frame of reference, and chemical potential, 
$\tilde{\mu}_{m}$, given by 
\begin{equation}
\Omega \tilde{\mu}_{m} 
= E_{m}^{sol} 
+ \sum_{n=1}^{M-1} E_{mn}^{int}c_n 
+ R \theta \ln \frac{c_m}{c_M} 
+ \alpha_m \left(c_m - \tilde{c}_m\right) 
- \Omega\nu_{m}\left(\transpose{\Fc}\sPK\right) 
\cdot 
\eyetwo.
\label{equ:chempotrel}
\end{equation}
On this basis, non-negativity of $M_m$ is sufficient to ensure non-negative dissipation due to component diffusion. 

\section{Numerical methods}
\label{sec: numerics}
The constitutive model introduced in Section 2 is implemented in the freeware material simulation kit, DAMASK \citep{Roters2019}, and a large-scale parallel finite element (FE) code using the PETSc numerical library \citep{Balay2015} is developed to handle the discretisation and numerical solution of the coupled field equations.

\subsection{Rate variational formulation of the initial boundary-value problems}

The following represents a special case of the variational treatment of combined Cahn-Hilliard and Allen-Cahn modelling of finite-deformation gradient elastic solids in \citet{Gladkov2015}. 

As detailed in \cite{Sve04}, a rate-variational-based formulation of the IBVPs is contingent in particular on the existence of a dissipation potential for the model in question. 
In the current case, the force-based form
\begin{equation}
d=\frac{1}{n+1}\sum\limits_{\alpha} g^\alpha\dot\gamma_0\abs{\frac{\tau^\alpha}{g^\alpha}}^{n+1} + \frac{1}{2}\sum\limits_{m=1}^{M-1}\nabla\tilde{\mu}_{m}\cdot M_m\nabla\tilde{\mu}_{m},
\label{equ:PotDisMecChe}
\end{equation} 
of this potential, $d$, determines the fluxes $\dot\gamma^\alpha = \partial_{\tau_{\alpha}}d$ and $\vctr{j}_{m} = -\partial_{\Grad\tilde{\mu}_{m}}d$, consistent with \eqref{equ:RulFloSysGli} and \eqref{equ:FluCom}, 
respectively. 
The convexity of \(d\) in the forces, and its non-negativity \(d\geqslant 0\), together imply $\delta\geqslant 0$ in the context of \eqref{equ:DenRatDisRes}.
Besides this potential, the rate-variational formulation is based on the energy storage-rate density, $\zeta$, and the supply rate density, $p_\mathrm{s}$. 
For the current model, this is given by 
\begin{equation}
\zeta := \sum\limits_{m=1}^{M-1}\left(\tilde{\mu}_{m}\dot{c}_{m} + \partial_{\tilde{c}_m}\psi\dot{\tilde{c}}_m\right ) + \partial_{\Grad \defmap}\psi\cdot\Grad\dot{\defmap} - \sum\limits_{\alpha}\tau_{\alpha}\dot{\gamma}_{\alpha}, \quad p_\mathrm{s} = \sum\limits_{m=1}^{M-1}\tilde{\mu}_{m}\sigma_{m}.
\label{equ:DenRatStoEneMix}
\end{equation} 
Together, these determine the volumetric part of the rate functional 
\begin{equation}
P = \int_{\domain{B}_{0}} p_{\mathrm{v}} \ \inc{\vctr{x}} + \int_{\partial\domain{B}_{0}^{\mathbf{f}}} p_{\mathrm{f}} \ \inc{\vctr{s}}
\,,\quad \text{where} \quad
p_{\mathrm{v}}
=\zeta+d+p_{\mathrm{s}}.
\label{equ:FunRat}
\end{equation}
The flux boundary conditions on $\partial\domain{B}_{0}^{\mathbf{f}}$ is given by $p_{\mathrm{f}}$.
The first variation of $P$ with respect to $\tilde{\mu}_{m}$, $\dot{\defmap}$, $\dot{\tilde{c}}_m$, and $\tau_\alpha$ takes the form 
\begin{align}
\delta P = & \sum\limits_{m=1}^{M-1} \int_{\domain{B}_{0}} \delta_{\tilde{\mu}_{m}} p_\mathrm{v} \delta\tilde{\mu}_{m}\ \inc{\vctr{x}} + \sum\limits_{m=1}^{M-1} \int_{\partial\domain{B}_{0}^{\mathbf{f}}}\left(\partial_{\Grad\tilde{\mu}_{m}} p_\mathrm{v}\,\bm{n}_{\partial \domain{B}_{0}} +\partial_{\tilde{\mu}_{m}} p_\mathrm{f}\right) \delta \tilde{\mu}_{m} \ \inc{\vctr{s}}\nonumber\\
+ & \int_{\domain{B}_{0}} \delta_{\dot\defmap} p_\mathrm{v} \delta\dot\defmap\ \inc{\vctr{x}} + \int_{\partial\domain{B}_{0}^{\mathbf{f}}} \left(\partial_{\Grad\dot\defmap} p_\mathrm{v}\,\bm{n}_{\partial \domain{B}_{0}} +\partial_{\dot\defmap} p_\mathrm{f}\right) \delta\dot\defmap\ \inc{\vctr{s}}\nonumber\\
+ & \sum\limits_{m=1}^{M-1}\int_{\domain{B}_{0}} \partial_{\dot{\tilde{c}}_m} p_{\mathrm{v}} \delta\dot{\tilde{c}}_m\ \inc{\vctr{x}} + \sum\limits_{\alpha}\int_{\domain{B}_{0}} \partial_{\tau_\alpha} p_{\mathrm{v}} \delta\tau_\alpha\ \inc{\vctr{x}}
\label{equ:VarPotRatMix}
\end{align} 
via integration by parts and the divergence theorem. 
Necessary for stationarity, $\delta P=0$, of \eqref{equ:VarPotRatMix} are the weak forms for the field relations
\begin{align}
\int_{\domain{B}_{0}} \partial_{\Grad \defmap} \psi \cdot \Grad \delta \dot \defmap\ \inc{\vctr{x}} - \int_{\partial \domain{B}_{0}}\partial_{\dot\defmap} p_{\mathrm{f}} \delta \dot \defmap\ \inc{\vctr{s}} = 0,\nonumber\\
\int_{\domain{B}_{0}} \left[\left(\dot{c}_m-\sigma\right) \delta\tilde{\mu}_{m} +M_m\Grad \tilde{\mu}_{m}\cdot\Grad\delta\tilde{\mu}_{m}\right]\ \inc{\vctr{x}} = 0,\nonumber\\
\int_{\domain{B}_{0}} \left[ \alpha_m \left({c}_m-\tilde{c}_m\right) \delta\tilde{c}_m +\kappa_m\Grad \tilde{c}_m\cdot\Grad\delta\tilde{c}_m\right]\ \inc{\vctr{x}} = 0,
\label{equ:RelFieWeaVarRat}
\end{align} 
and constitutive relation $0 = - \dot{\gamma}^\alpha + \partial_{\tau_\alpha} d$.

\subsection{Finite element implementation}

The rate-variational formulation yields the weak form of the field relations required for their finite-element (FE) implementation. 
In particular, \cref{equ:RelFieWeaVarRat} yields directly the weak momentum balance relation
\begin{equation}
\vctr{0} = \int_{\mathcal{B}_{0}} \nabla\delta\dot{\defmap} \cdot \fPK \inc{\vctr{x}},
\label{equ:BalMomWea}
\end{equation}
where $p_{\mathrm{f}} = 0$ is assumed for simplicity, the weak multi-component transport relation 
\begin{align}
\int_{\domain{B}_{0}} \left[\left(\dot{c}_m-\sigma\right) \delta\tilde{\mu}_{m} +M_m\Grad \tilde{\mu}_{m}\cdot\Grad\delta\tilde{\mu}_{m}\right]\ \inc{\vctr{x}} = 0,
\label{equ:BalCompWea}
\end{align}
and the weak non-local relation
\begin{align}
\int_{\domain{B}_{0}} \left[ \alpha_m \left({c}_m-\tilde{c}_m\right) \delta\tilde{c}_m +\kappa_m\Grad \tilde{c}_m\cdot\Grad\delta\tilde{c}_m\right]\ \inc{\vctr{x}} = 0.
\label{equ:BalNLWea}
\end{align}
where $\delta \dot\defmap$, $\delta \tilde{\mu}_{m}$ and $\delta \dot{\tilde c}_m$ are the virtual deformation rate, chemical potential and non-local concentration fields respectively. 
No-flux boundary conditions are assumed for the sake of simplicity.

The deformation field, $\defmap\,(\x)$, chemical potential, $\tilde{\mu}_{m}\,(\x)$, and non-local concentration field, $\tilde{c}_m\,(\x)$, in addition to their virtual counterparts are discretised using a FE basis of shape functions, $N_{i}^{\defmap}$, $N_{i}^{\mu}$, $N_{i}^{\tilde{c}}$, $N_{i}^{\delta \dot\defmap}$, $N_{i}^{\delta \mu}$, and $N_{i}^{\delta \tilde{c}}$, where $[\defmap]_{i}$, $[\tilde{\mu}_{m}]_i$, $[\tilde{c}_m]_i$, ${[\delta \dot\defmap]_{i}}$, ${[\delta\tilde{\mu}_{m}]_i}$, and $[\delta \tilde{c}_{m}]_i$ are the respective degrees of freedom.
The corresponding discrete differential operator matrices are $ \tnsr B_{i}^{\defmap}$, $\tnsr B_{i}^{\mu}$, $\tnsr B_{i}^{\tilde{c}}$, $\tnsr B_{i}^{\delta \dot\defmap}$, $\tnsr B_{i}^{\delta \mu}$ and $\tnsr B_{i}^{\delta \tilde{c}}$. 
Under these approximations, the weak forms \cref{equ:BalMomWea,equ:BalCompWea,equ:BalNLWea} can be rewritten as 
\begin{equation}
\label{eqn:weak-mechD}
 \sum_{i} [\delta \dot\defmap]_{i}^{T} \underbrace{\int_{\domain B_{0}}  [\tnsr B_{i}^{\delta \dot\defmap}]^{T} {\fPK} \inc{\vctr{x}}}_{\mathcal{R}_{i}^\text{mech}} = \vctr{0},
\end{equation}
\begin{align}
\label{eqn:weak-compD}
 \sum_{i} [\delta\tilde{\mu}_{m}]_i^{T} \underbrace{\int_{\domain B_{0}}  \left[ [N_{i}^{\delta \mu}]^{T}\left(\dot{c}_m - \sigma_m\right) + [\tnsr B_{i}^{\delta \mu}]^{T} {M}_m \tnsr B_{i}^{\mu}[\tilde{\mu}_{m}]_i \right] \inc{\vctr{x}}}_{\mathcal{R}_{i,m}^\text{chem}} = 0,
 \end{align}
\begin{align}
\label{eqn:weak-NLD}
 \sum_{i} [\delta \tilde{c}_{m}]_i^{T} \underbrace{\int_{\domain B_{0}}  \left[ [N_{i}^{\delta \tilde{c}}]^{T}\alpha_m\left(c_m - N_{i}^{\tilde{c}} [\tilde{c}_m]_i\right) + [\tnsr B_{i}^{\delta \tilde{c}}]^{T} \kappa_m \tnsr B_{i}^{\tilde{c}}[\tilde{c}_{m}]_i \right] \inc{\vctr{x}}}_{\mathcal{R}_{i,m}^\text{NL}} = 0,
 \end{align}
which defines a non-linear system of equations for the unknowns $[\defmap]_{i}$, $[\tilde{\mu}_{m}]_i$, and $[\tilde{c}_m]_i$.
A time-discrete system of equations is obtained by using a backward \Euler\ approximation 
\begin{equation}
\dot{c}_m = \frac{c_m(t_{n}) - c_m(t_{n-1})}{\triangle t}
\end{equation}
of the rate $\dot{c}_m$ in \cref{eqn:weak-compD}.

The solution approach followed in this work involves solving the coupled system of \cref{eqn:weak-mechD,eqn:weak-compD,eqn:weak-NLD} within a staggered iterative loop until a self consistent solution is achieved for a time increment.

\subsection{Chemical potential solution}

The solution of the coupled system of \cref{eqn:weak-mechD,eqn:weak-compD,eqn:weak-NLD}, for the deformation field, $\defmap$, chemical potential, $\tilde{\mu}_{m}$, and non-local concentration field, $\tilde{c}_m$, requires the inversion of \cref{equ:chempotrel} in order to express $c_m := c_m ({\tilde{\mu}_{n}})$ for $m,n=1, \ldots, M-1$.
This is achieved algorithmically in the current work through a semi-implicit splitting of the chemical potential relation,
\begin{equation}
\tilde{\mu}_{m}(t_{n}) = \check{\mu}_{\tilde{m}}(t_{n}) + \hat{\mu}_{\tilde{m}}(t_{n-1}), 
\label{equ:split1}
\end{equation}
into a convex
\begin{align}
\Omega\check{\mu}_{\tilde{m}}(t_{n}) &= E_{m}^{sol} + R \theta \ln \frac{c_m(t_{n})}{c_M(t_{n})}+ \alpha_m \left(c_m(t_{n}) - \tilde{c}_m)\right) - \Omega\nu_{m}\left(\transpose{\Fc}\sPK\right) \cdot \eyetwo, 
\label{equ:split2}
\end{align}
and non-convex, \ie,
\begin{equation}
\Omega\hat{\mu}_{\tilde{m}}(t_{n-1}) = \sum_{n=1}^{M-1} E_{mn}^{int}c_n(t_{n-1}),
\label{equ:split3}
\end{equation}
contribution.
\Cref{equ:split2} can then be inverted to express $c_m(t_{n})$ in terms of \cref{equ:split1,equ:split3}, yielding,
\begin{equation}
c_m(t_{n}) = \frac{\exp{\left(\frac{f_m - \alpha_m c_m(t_{n})}{R\theta}\right)}}{1 + \sum_{n=1}^{M-1}\exp{\left(\frac{f_n - \alpha_n c_n(t_{n})}{R\theta}\right)}},
\label{equ:split4}
\end{equation}
where
\begin{equation}
f_m =  \Omega\tilde{\mu}_{m}(t_{n}) - E_{m}^{sol} + \alpha_m \tilde{c}_m + \Omega\nu_{m}\left(\transpose{\Fc}\sPK\right) \cdot \eyetwo.
\label{equ:split5}
\end{equation}
\Cref{equ:split4} is an implicit system of equations to be numerically solved for $c_m(t_{n})$ for a given set ${\tilde{\mu}_{n}}(t_{n})$. 
A fixed point iteration is employed, which is unconditionally convergent for $\alpha_m \ge 0$ as the the fixed point operator, \ie\ the RHS in \cref{equ:split4}, is guaranteed to be a contractive mapping.

\section{Results and discussion}
In this section, the developed chemo-mechanical model for multi-component finite-strain elasto-viscoplastic materials is validated, benchmarked and showcased through illustrative examples. 
\subsection{Validation and benchmarking of the numerical scheme}
\label{sec: convergence}
Diffusion simulations are first performed to study the convergence, accuracy and performance of the proposed numerical scheme for the CH model.
\subsubsection{Convergence behaviour}
\label{sec: penaltyAndgradient}
\begin{figure}
	\centering
	\begin{tabular}{lcrl}
		\begin{overpic}[width=0.65\textwidth]{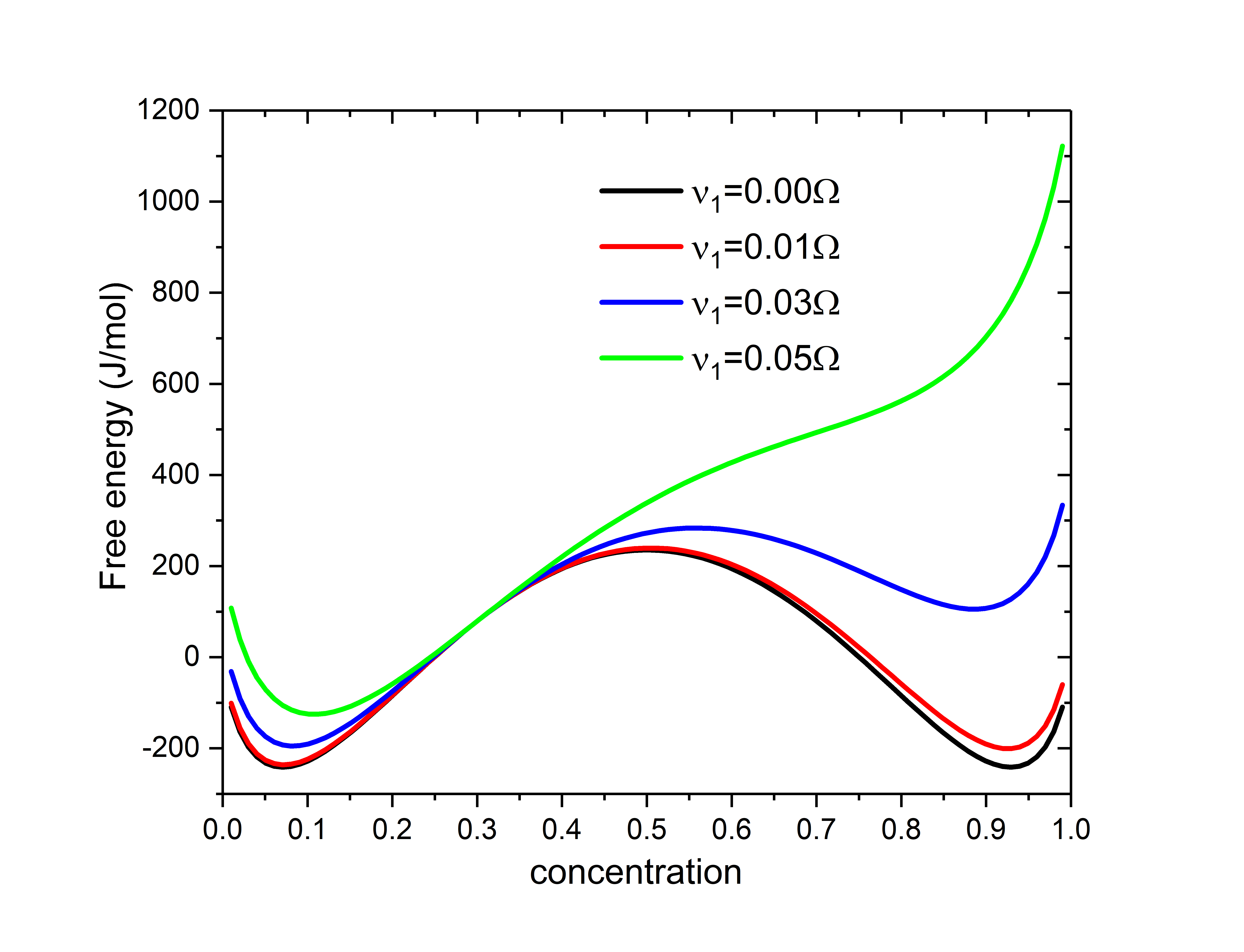}\end{overpic}
	\end{tabular}
	\caption{Influence of the solute volumetric mismatch coefficient, $\nu_{1}$, on the free energy, $\psi$, for $\nu_{1} = 0.01, 0.03, \ \text{and} \ 0.05$.
	      	It is assumed that the volumetric mismatch between solute components is accommodated by elastic deformation.
	}
	\label{fig: energyplot}
\end{figure}

A non-local concentration field, $\tilde{c}$, is introduced in the current model to account for the gradient energy contributions, thereby reducing the strong fourth-order CH PDE to two weakly non-local second-order PDEs.
The conditions which should be fulfilled for the proposed method to approach the results by the strong CH solution are discussed in this section.
To investigate the effect of the penalty parameter, $\alpha$, and the gradient coefficient, $\kappa$, on the equilibrium numerical solution, we study one-dimensional (1-D) diffusion simulations without mechanical deformation.
Parameters for the chemical free energy density function in \cref{equ:DenEneFreCheDamUn} are taken as 
$\Omega = 1\times10^{-5} \text{m}^3 \text{mol}^{-1}, E_1^{sol}=1.24\times10^{4} \text{J} \text{mol}^{-1}, E_{11}^{int}=-1.24\times10^{4} \text{J} \text{mol}^{-1}, \theta=498\text{K}, \kappa_1=1\times10^{-16}\text{J} \text{m}^{2}\text{mol}^{-1}, M_1 = 2.2\times10^{-19}\text{m}^{5} \text{s}^{-1} \text{J}^{-1}$ and the penalty parameter $\alpha_1$ is varied from $4\times10^{3}$ to $2.5\times10^{6} \text{J} \text{mol}^{-1}$.
The resulting spinodal compositions are $c_\alpha=0.07$ and $c_\beta=0.93$, respectively.
The chemical free energy without mechanical deformation of the studied system is represented by the black curve in \cref{fig: energyplot}.
The minima in the double-well free energy curve represent the spinodal compositions resulting from the decomposition of an initial homogeneous mixture.

The 1-D domain,  $\domain B_0 = [0,2L]$, is initially into two regions, $\domain B_\alpha = [0,L)$ and $\domain B_\beta = (L,1.0]$.
The initial concentrations for these two regions are taken as $c_\alpha = 0.49$ and $c_\beta = 0.51$, respectively.
The system size, $2L$, is \SI{3}{\micro \meter} and meshed with $300$ regular hexahedral elements.

\begin{figure}
	\centering
	\begin{tabular}{lcrl}
		\begin{overpic}[width=0.9\textwidth]{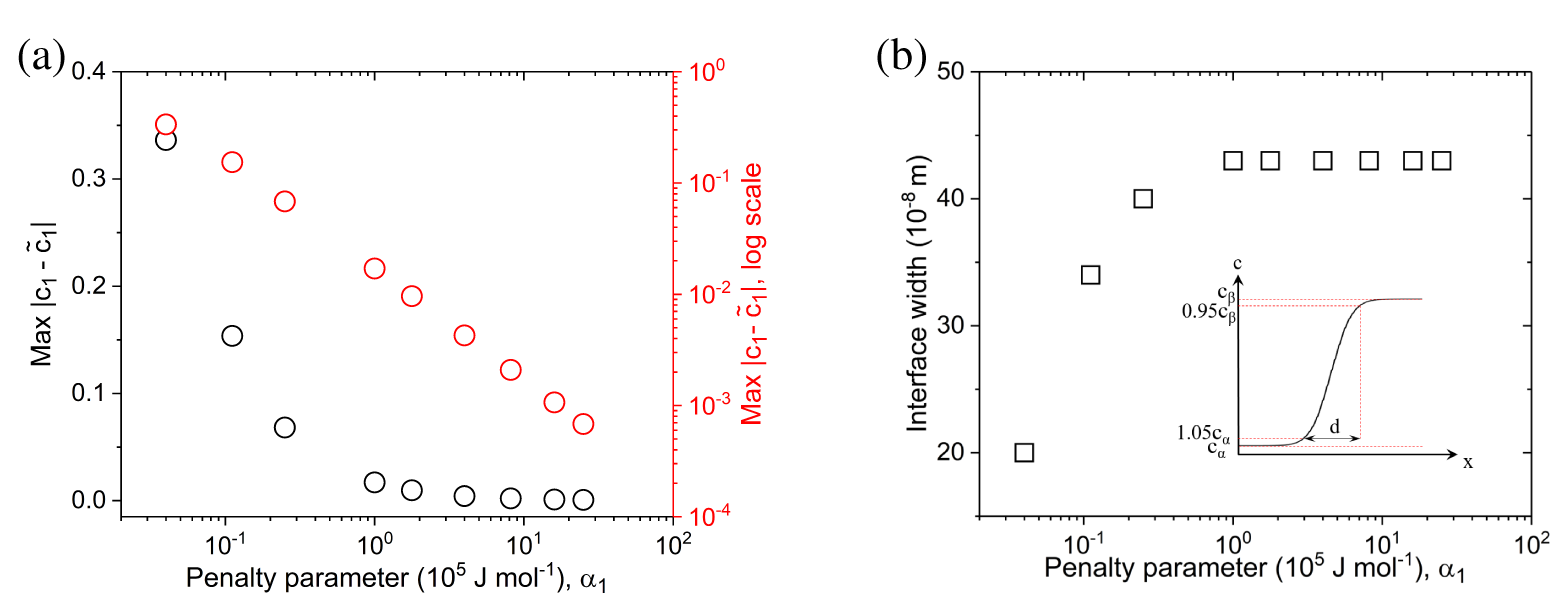}\end{overpic}
	\end{tabular}
	\caption{(a) Convergence of the concentration, $c$, to the non-local concentration, $\tilde{c}$, as a function of the penalty parameter, $\alpha_1$, ranging from $4\times10^{3}$ to $2.5\times10^{6} \text{J} \text{mol}^{-1}$, in linear (black) and log (red) scale.
		(b) Relationship between the interface width and the penalty parameter at a constant gradient coefficient $\kappa_1=1\times10^{-16} \text{J} \text{m}^{2}\text{mol}^{-1}$.
		Schematic of the interface between binodal concentration $c_{\alpha}$ and $c_{\beta}$ is given in the inset.
		The interface width $d$ is defined as the distance where the concentration changes from $1.05c_{\alpha}$ to $0.95c_{\beta}$.
	}
	\label{fig: Penalty-Conc-width}
\end{figure}

\Cref{fig: Penalty-Conc-width}(a) shows the maximum of the pointwise difference between concentration and non-local concentration, $\text{max}_{x \in [0,2L]}|c(x)-\tilde{c}(x)|$, at steady state, with increasing values of the penalty parameter.  
The difference decreases substantially from $3.4\times10^{-1}$ to $6.8\times10^{-4}$ with increasing the penalty parameter from $4.0\times10^{3}$ to $2.5\times10^{6} \text{J} \text{mol}^{-1}$.
Convergence of the concentration, $c$, to the non-local concentration, $\tilde{c}$, is observed with an exponential reduction in the difference on increasing the value of the penalty parameter, $\alpha$.
\Cref{fig: Penalty-Conc-width}(b) presents the variation of the interface width, $d$, with increasing the penalty parameter. 
The interface width converges to a constant value of $4.3\times10^{-7} m$ with minimal differences when the penalty parameter is larger than $1.0\times10^{5} \text{J} \text{mol}^{-1}$.
The convergence of the interface width in such studies is a good indicator for choosing an appropriate penalty parameter.

\begin{figure}
	\centering
	\begin{tabular}{lcrl}
		\begin{overpic}[width=0.65\textwidth]{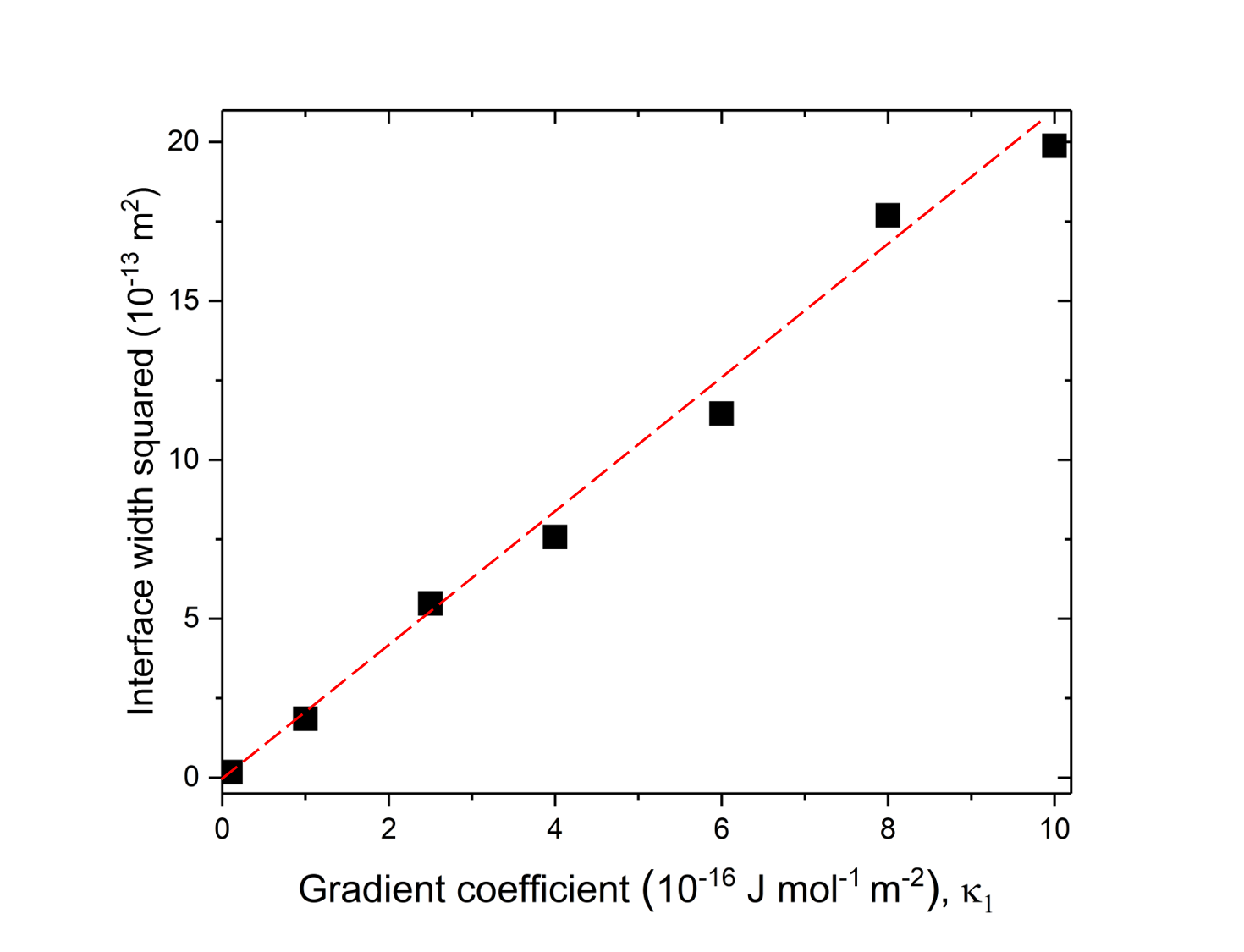}\end{overpic}
	\end{tabular}
	\caption{Interface width squared, $l^2$, as a function of the gradient coefficient, $\kappa_1$ varrying from $1\times10^{-17}$ to $1\times10^{-15} \text{J} \text{m}^{2}\text{mol}^{-1}$, obtained from simulations (squares) and compared with the ideal Cahn-Hilliard scaling (line).
	}
	\label{fig: GraIntWidth}
\end{figure}

An alternative way to characterize the convergence of the proposed numerical algorithm is to study the relationship between the gradient coefficient, $\kappa_m$, and the interface width, $d$.
Based on the classical CH theory \citep{Cahn1958}, the gradient coefficient should have quadratic relationship with the interface width.
\Cref{fig: GraIntWidth} shows the simulated interface width squared as a function of the gradient coefficient at steady state. 
The gradient coefficient $\kappa_1$ varies from $1\times10^{-17}$ to $1\times10^{-15}\text{J} \text{m}^{2}\text{mol}^{-1}$.
Note that for each simulation, the penalty parameter is chosen large enough to guarantee that the interface width converges to the constant value as discussed above.
\Cref{fig: GraIntWidth} verifies the linear relationship between the interface width and the gradient coefficient, and shows that the penalty parameter does not affect the simulation results when it is sufficiently large.
The above results indicate that the numerical solutions of the proposed approach converges to the conventional CH model for a sufficiently large penalty parameter, and guidelines for choosing the penalty parameter are provided.

\subsubsection{Numerical performance}
\label{sec: comparision_conc_and_mu}
In this section, the accuracy and performance of the proposed numerical approach is compared to a conventional CH solution scheme.
A two-dimensional spinodal decomposition example in the absence of mechanical deformation is solved to compare the two solution schemes.
The model parameters used are as follows: $\Omega = 1\times10^{-5} \text{m}^3 \text{mol}^{-1}, E_1^{sol}=2\times10^{4} \text{J} \text{mol}^{-1}, E_{11}^{int}=-5\times10^{4} \text{J} \text{mol}^{-1}, R\theta=0.1\times10^{4} \text{J} \text{mol}^{-1}, \kappa_1=1\times10^{-16}\text{J}\text{m}^{2}\text{mol}^{-1}, M_m = 1.0\times10^{-19}\text{m}^{5} \text{s}^{-1} \text{J}^{-1}$, $\alpha_1 = 5\times10^{5} \text{J} \text{mol}^{-1}$.
A square domain, $\domain B_0 = [0,L] \times [0,L]$, of size $L = \SI{2.56}{\micro \meter}$ is meshed with $256\times256$ regular hexahedral elements.
The initial concentration is taken as 0.5 with a random perturbation of amplitude 0.1.

\begin{figure}
	\centering
	\begin{tabular}{lcrl}
		\begin{overpic}[width=0.9\textwidth]{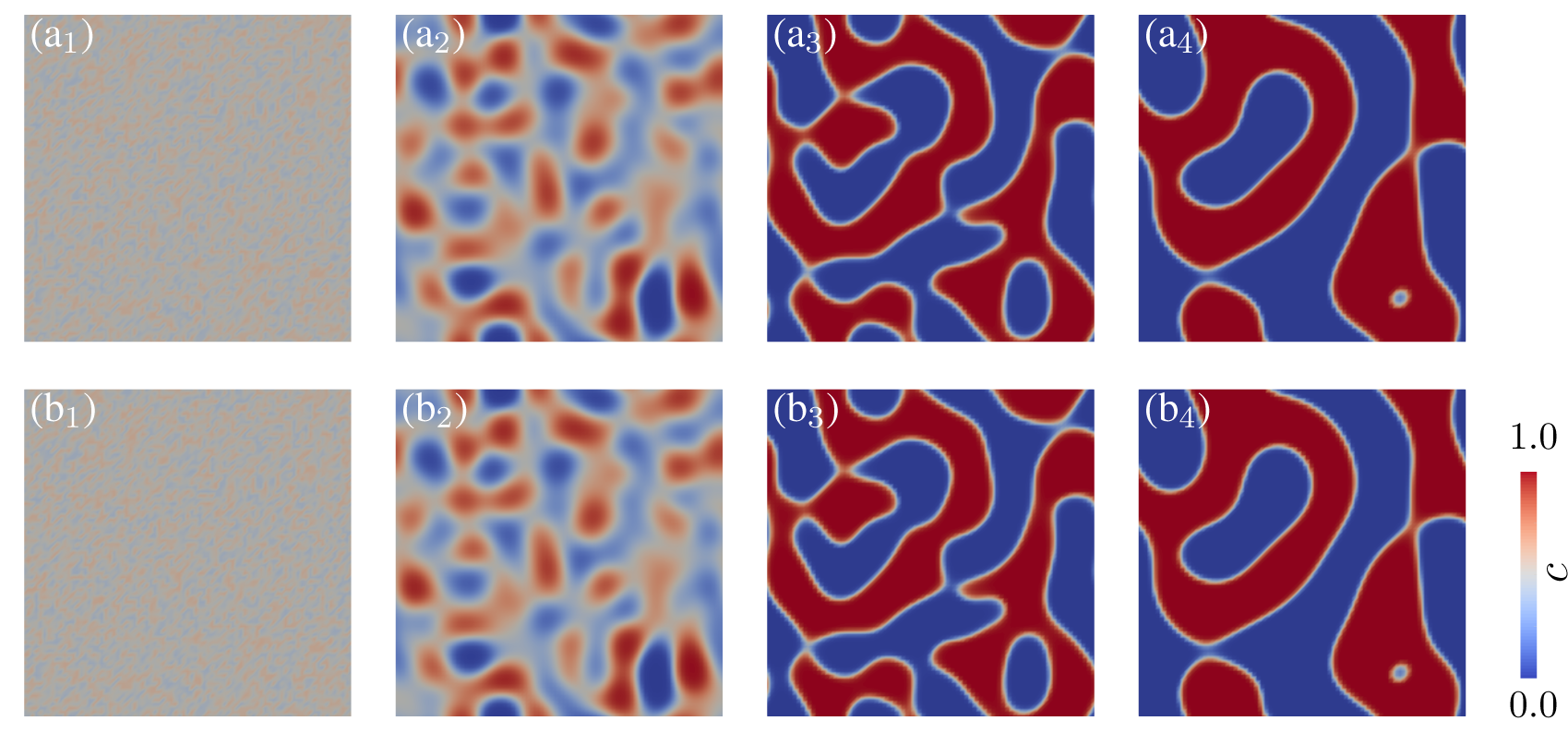}\end{overpic}
	\end{tabular}
	\caption{%
		Spinodal decomposition of a homogeneous solid solution with initial concentration of 0.5 and random perturbation of amplitude 0.1 at (1) 1s, (2) 500s, (3) 1000s, and (4) 1800s.
		The numerical implementations of the transport relations is based on (a) the concentration field and (b) the chemical potential field, respectively.
	}
	\label{fig: conc_mu_compa}
\end{figure}

\begin{figure}
	\centering
	\begin{tabular}{lcrl}
		\begin{overpic}[width=0.65\textwidth]{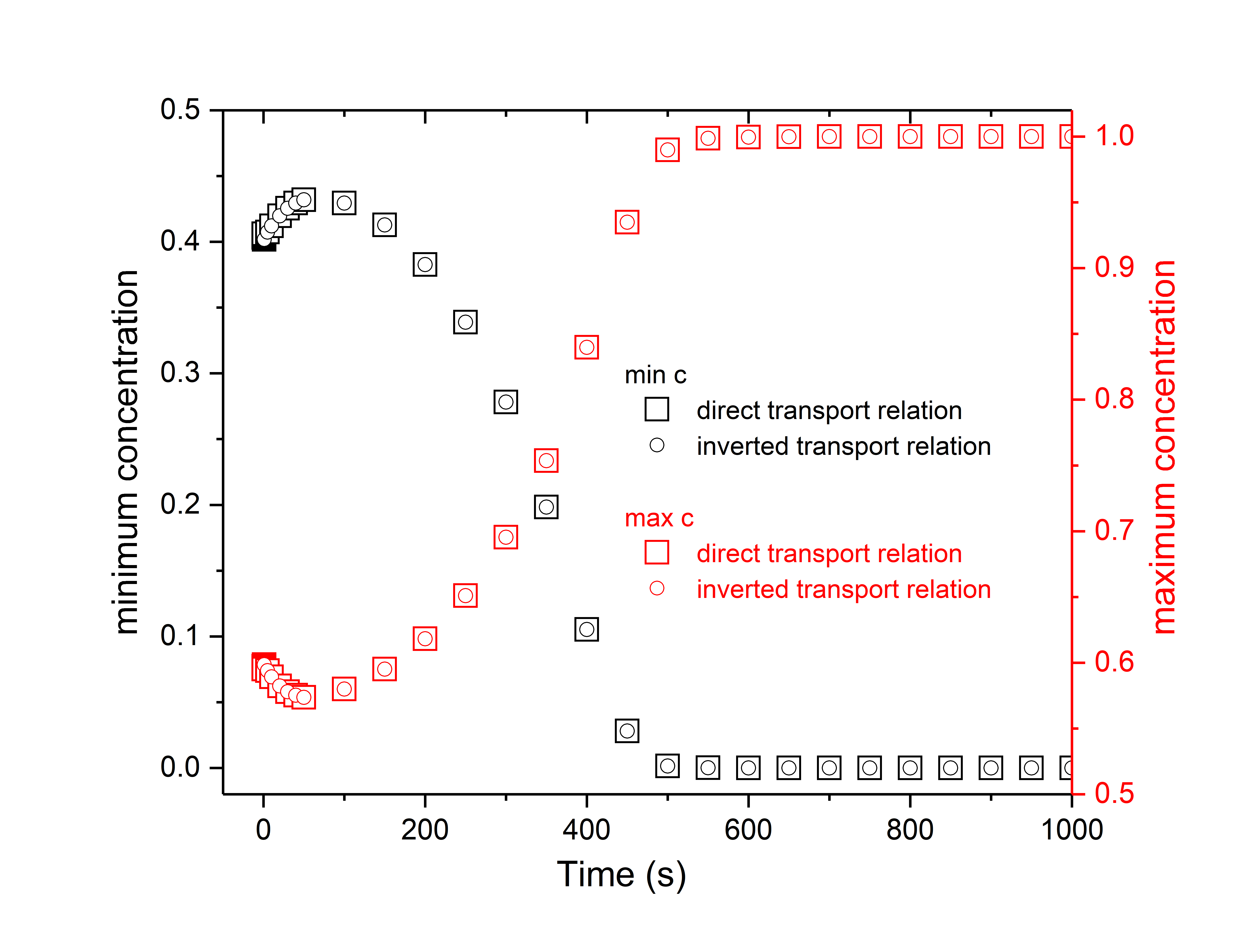}\end{overpic}
	\end{tabular}
	\caption{ Evolution of the maximum (red) and minimum (black) values of the concentration during the spinodal decomposition process obtained from classical (squares) and proposed (circles) numerical approaches, respectively.
	}
	\label{fig: max_min_conc}
\end{figure}

\Cref{fig: conc_mu_compa} shows the temporal evolution of the concentration field during spinodal decomposition, starting from a homogeneous mixture with an average concentration of 0.5 and random perturbation with an amplitude of 0.1, into two spinodal compositions of $1.0\times10^{-4}$ and $9.9\times10^{-1}$, determined by the minima of the free energy, $\psi$. 
Negligible differences in the concentration fields between the two solution approaches are observed.
In order to quantitatively compare the simulation results, 
\cref{fig: max_min_conc} shows the evolution of the maximum and minimum values of the concentration during the spinodal decomposition process, with spinodal decomposition initiating at 500s.
It can be seen that the simulated results for these two different approaches completely overlap during the entire spinodal decomposition process, thus validating the accuracy of the proposed numerical approach.

\begin{figure}
	\centering
	\begin{tabular}{lcrl}
		\begin{overpic}[width=0.65\textwidth]{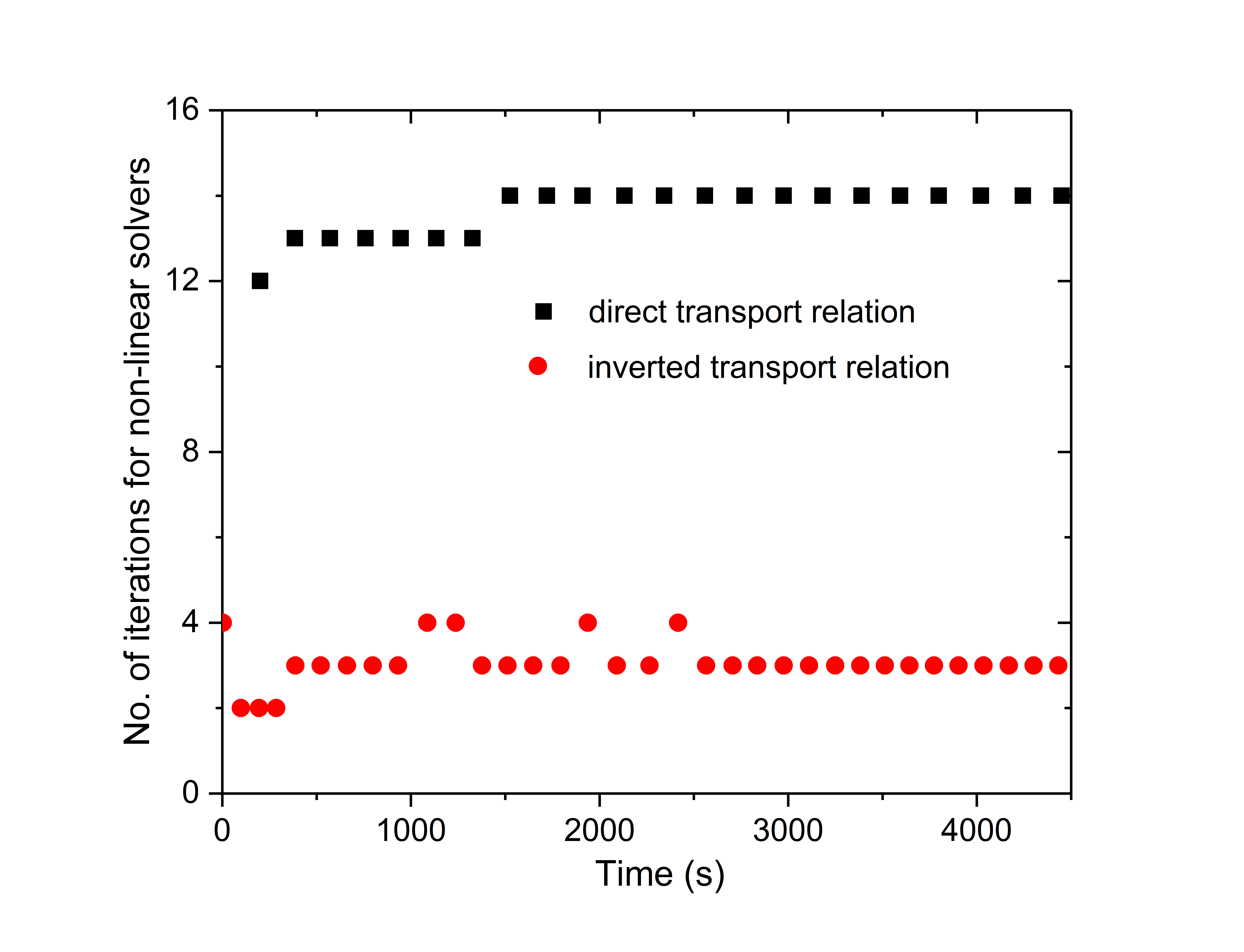}\end{overpic}
	\end{tabular}
	\caption{Number of Newton iterations required to solve each time step during the simulation process for the direct (black squares) and inverted (red circles) transport relations.
	}
	\label{fig: NoofIterations}
\end{figure}

In order to compare the numerical performance of the two numerical approaches, the same Newton-Raphson scheme is used in the above simulations.
The number of iterations required to obtain the solution for the nonlinear Newton-Raphson solver at each  time step is used to quantify the performance and efficiency of the numerical scheme.
The number of Newton iterations for both approaches during the calculations is shown in \cref{fig: NoofIterations}.
The number of Newton iterations for solving the inverted transport relations (2-4 iterations) is significantly lower than that for solving the conventional transport relations (12-14 iterations).
These comparisons demonstrate the performance and the efficiency of the proposed numerical scheme.

\subsection{Chemo-mechanical coupling}
\label{sec: two-dimensional}
In this section, the chemical diffusion model is coupled to a finite-strain mechanical model to study the role of mechanical deformation on the spinodal decomposition process.
Both elastic and plastic deformation accompanying diffusion are considered in the simulations. 
The material parameters for the chemical energy and crystal plasticity model are listed in \cref{Tab: matepara2D}.
To illustrate the effect of mechanical deformation, through the deformation arising from volumetric mismatch between solute components, $\Fc$ in \cref{eq:inelastic flow rule}, on the free energy of the system, \cref{fig: energyplot} shows the free energy of a 1-D system with various values of the solute volumetric mismatch coefficient.
Note that in this plot, it is assumed that the local deformation arising from volumetric mismatch between solute components can only be accommodated by elastic deformation.
It can be seen that in terms of kinetics, the driving force for spinodal decomposition decreases with increasing levels of volumetric mismatch. 
From an energetic point of view, the miscibility gap is also modified, with the solute-poor spinodal being more energetically favoured.

\begin{table}[h!]
	\centering
	\caption{Chemo-mechanical material parameters for the binary spinodal system. $\Omega$ is the molar volume, $E_1^{sol}$ is the component solution energy, $E_{11}^{int}$ is the component interaction energy, $\theta$ is the temperature, $\kappa_1$ is the gradient coefficient, $\alpha_1$ is the penalty parameter, $M_1$ is the component mobility, $\dot{\gamma}_0$ is the reference shear rate, $n$ is the strain-rate sensitivity exponent, $g_0$ and $g_\infty$ are the initial and asymptotic slip resistances, and $h_0$ and $h_{\alpha \beta}$ are hardening parameters.}
	\resizebox{0.9\linewidth}{!}{%
		\begin{tabular}{ccccccccccc}
			\toprule
			\textbf{Chemical energy} & $\Omega$ ($\text{m}^3 \text{mol}^{-1}$) & $E_1^{sol}$ ($\text{J}\text{mol}^{-1}$) & $E_{11}^{int}$ ($\text{J}\text{mol}^{-1}$) & $\theta$ (K) \\
			\midrule		
			& $1\times10^{-5}$ & $1.24\times10^{4}$ & $-1.24\times10^{4}$ & 498  \\
			\midrule
			& $\kappa_1$ ($\text{J} \text{m}^{2}\text{mol}^{-1}$) &$\alpha_1$ ($\text{J} \text{mol}^{-1}$) & $M_1$ ($\text{m}^{5} \text{s}^{-1} \text{J}^{-1}$)\\
			\midrule
			& $2.5\times10^{-17}$ &$5\times10^{5}$ & $2.2\times10^{-19}$ \\
			\midrule
			\textbf{Crystal plasticity model} &$\dot{\gamma}_0$ ($s^{-1}$) & n & $g_0$ ($\text{MPa}$) & $g_\infty$ ($\text{MPa}$) & a  \\
			\midrule
			&$1\times10^{-3}$  & 20  & 31 & 63  & 2.25\\
			\midrule
			& $h_0$ ($\text{MPa}$)& \text{Coplanar} $h_{\alpha \beta}$ ($\text{MPa}$) &\text{Non-coplanar} $h_{\alpha \beta}$ ($\text{MPa}$)\\
			\midrule
		    & 75 & 1 & 1.4\\
			\midrule
			\textbf{Elastic constants}& $C_{11}(Pa)$              & $C_{12}(Pa)$        & $C_{44}(Pa)$\\
			\midrule
			& $1.06\times10^{11}$ & $6.0\times10^{10}$ & $2.8\times10^{10}$ \\     
			\bottomrule
	\end{tabular}}
	\label{Tab: matepara2D}
\end{table}

\begin{figure}
	\centering
	\begin{tabular}{lcrl}
		\begin{overpic}[width=0.9\textwidth]{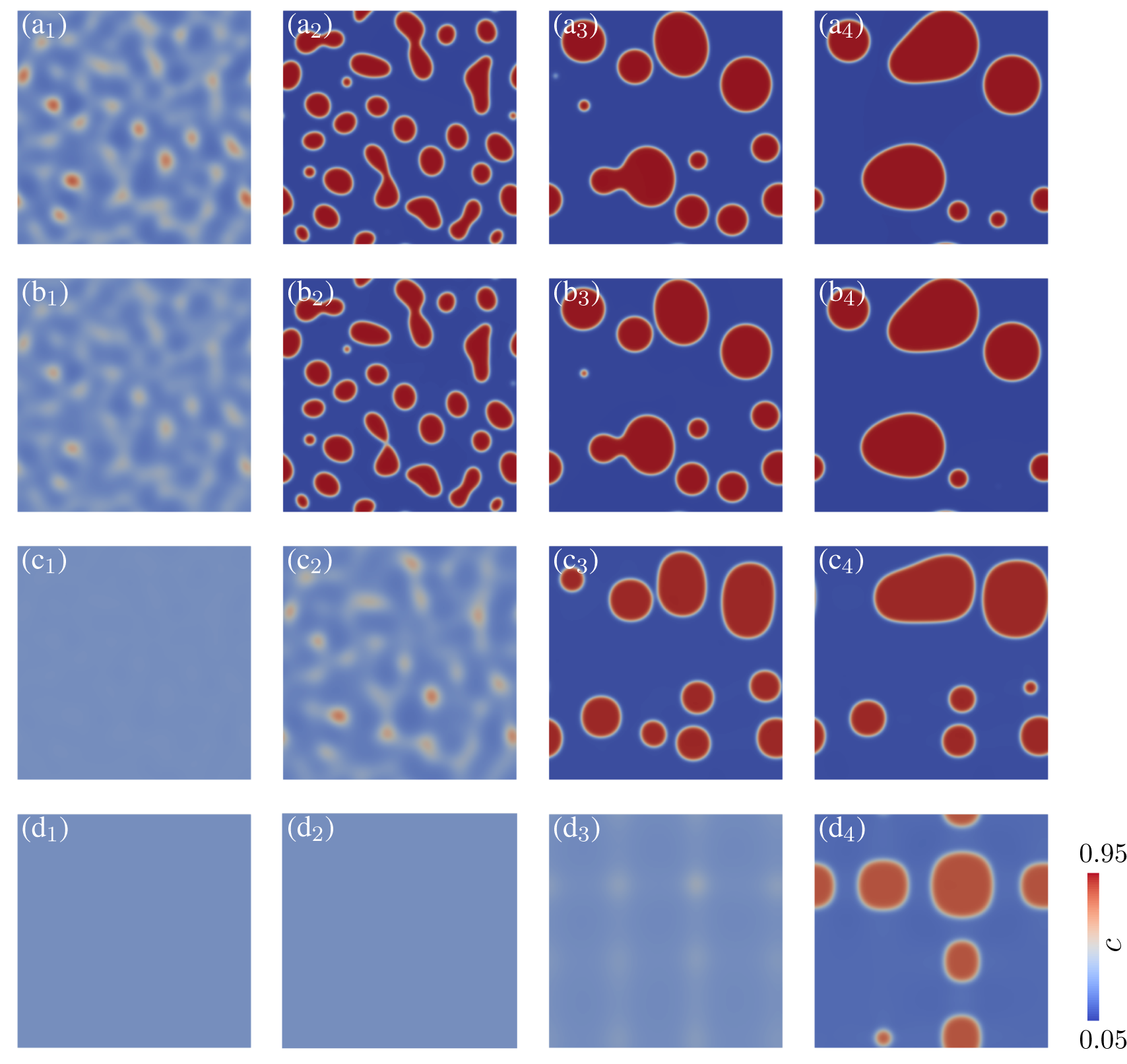}\end{overpic}
	\end{tabular}
	\caption{%
		Spinodal decomposition and coarsening (a) without volumetric mismatch between solute components, and with elastic accommodation of the volumetric mismatch between solute components, for mismatch strain coefficients of (b) $\nu_{1} =0.01$, (c) $\nu_{1} =0.03$, and (d) $\nu_{1} =0.05$, at time (1) 500s, (2) 800s, (3) 6000s, and (4) 10000s.
	}
	\label{fig: 2D_chem_elas}
\end{figure}

A square domain, $\domain B_0 = [0,L] \times [0,L]$, of size $L = \SI{2.56}{\micro \meter}$ is meshed with $256\times256$ regular hexahedral elements.
Periodic boundary conditions on both concentrations and displacements were applied.
The temporal evolution of the concentration field during the spinodal decomposition process, in the absence of mechanical deformation (\ie\ setting $\nu_{1} =0.00$), is shown in \cref{fig: 2D_chem_elas} (a).
At the early stage of the spinodal decomposition, the initial homogeneous mixture decomposes into two regions with different compositions, following their corresponding composition branch on the generalized muti-well Landau energy landscape \cref{fig: 2D_chem_elas} ($\text{a}_\text{1}$). 
The decomposition stage is driven by the minimization of the chemical free energy.
Following decomposition, coarsening can be observed from \cref{fig: 2D_chem_elas} ($\text{a}_\text{2}$) to ($\text{a}_\text{4}$), which is driven by the minimization of interface energy, that occurs at a longer time scale compared to the initial decomposition stage.

The effect of the volumetric mismatch between solute components on spinodal decomposition and coarsening is investigated by first considering the case of an elastically deforming material.
The volumetric mismatch coefficient, $\nu_{1}$, is varied between $0.01\ \text{and}\ 0.05$, and the material parameters used are listed in \cref{Tab: matepara2D}.
The temporal evolution of the concentration field during spinodal decomposition is shown in \cref{fig: 2D_chem_elas} (b) to (d).
It can be seen that the decomposition is minimally affected when the solute induced deformation is relatively small ($\ie, \nu_{1} = 0.01$), comparing \cref{fig: 2D_chem_elas} (a) and (b).
However, as the volumetric mismatch coefficient increases, the spinodal decomposition kinetics is significantly reduced, as shown in \cref{fig: 2D_chem_elas} (c) and (d).
This is due to the increasing elastic energy contribution to the driving force, which has the effect of suppressing the spinodal decomposition (\cref{fig: energyplot}).
The spinodal compositions are also affected by the chemo-mechanical coupling, with the solute-rich spinodal point decreasing from 0.93 to 0.82 when the volumetric mismatch coefficient increases up to 0.05.
Furthermore, one can observe that the morphology of the solute-rich region is also affected by the deformation arising from volumetric mismatch between solute components, comparing \cref{fig: 2D_chem_elas} (a) and (d).
In the absence of mechanical coupling, the solute-rich regions exhibit a spherical morphology due to the isotropic nature of the interface energy (\cref{fig: 2D_chem_elas} ($\text{a}_\text{4}$)), but with increasing volumetric mismatch between solute components, the solute-rich regions are observed to order themselves along the softer directions of the cubically anisotropic elastic stiffness tensor used here (\cref{fig: 2D_chem_elas} ($\text{d}_\text{4}$)).

To our knowledge, the numerical investigations of spinodal decomposition, presented in the literature, are limited to the coupling of elastic deformation with diffusion.
The capability of the developed chemo-mechanical model allows us to explore the more challenging influence of elasto-plasticity on the spinodal decomposition and coarsening, which is observed in materials applications such as rafting in superalloys \citep{Kontis2018} and hydride formation \citep{Korbmacher2018}.
Parameters used for the crystal plasticity model are listed in \cref{Tab: matepara2D}.
 
\begin{figure}
	\centering
	\begin{tabular}{lcrl}
		\begin{overpic}[width=0.9\textwidth]{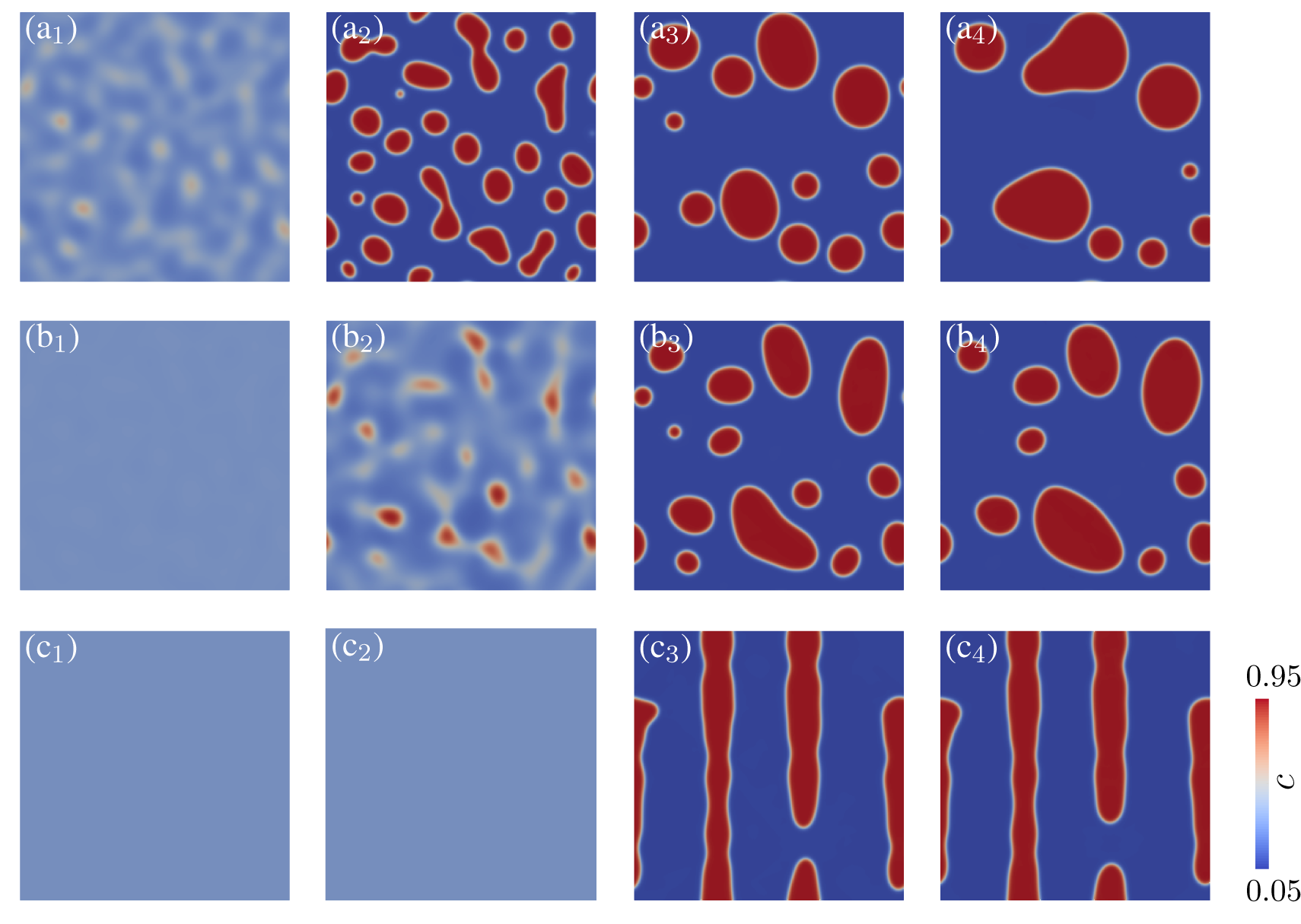}\end{overpic}
	\end{tabular}
	\caption{%
		Spinodal decomposition and coarsening with elasto-viscoplastic accommodation of the deformation due to the volumetric mismatch between solute components, for mismatch strain coefficients of, (a) $\nu_{1} =0.01$, (b) $\nu_{1} =0.03$, and (c) $\nu_{1} =0.05$, at time (1) 500s, (2) 800s, (3) 6000s, and (4) 10000s.
	}
	\label{fig: 2D_chem_plas}
\end{figure}

The temporal evolution of the concentration field during spinodal decomposition on an elastic-plastically deforming material is shown in \cref{fig: 2D_chem_plas}.
Comparing \cref{fig: 2D_chem_elas} and \cref{fig: 2D_chem_plas} reveals that visco-plastic relaxation significantly affects the decomposition process. 
Spinodal decomposition is observed at 6000s in the plastic material (\cref{fig: 2D_chem_plas} ($\text{c}_\text{3}$)) compared to 10000s in the elastic material (\cref{fig: 2D_chem_elas} $\text{a}_\text{4}$), for the largest volumetric mismatch coefficient considered. 
In addition, the spinodal compositions are close to those obtained from the simulations in the absence of mechanical deformation,
since the deformation arising from volumetric mismatch between solute components can be effectively relaxed by the plastic deformation.

\begin{figure}
	\centering
	\begin{tabular}{lcrl}
		\begin{overpic}[width=0.9\textwidth]{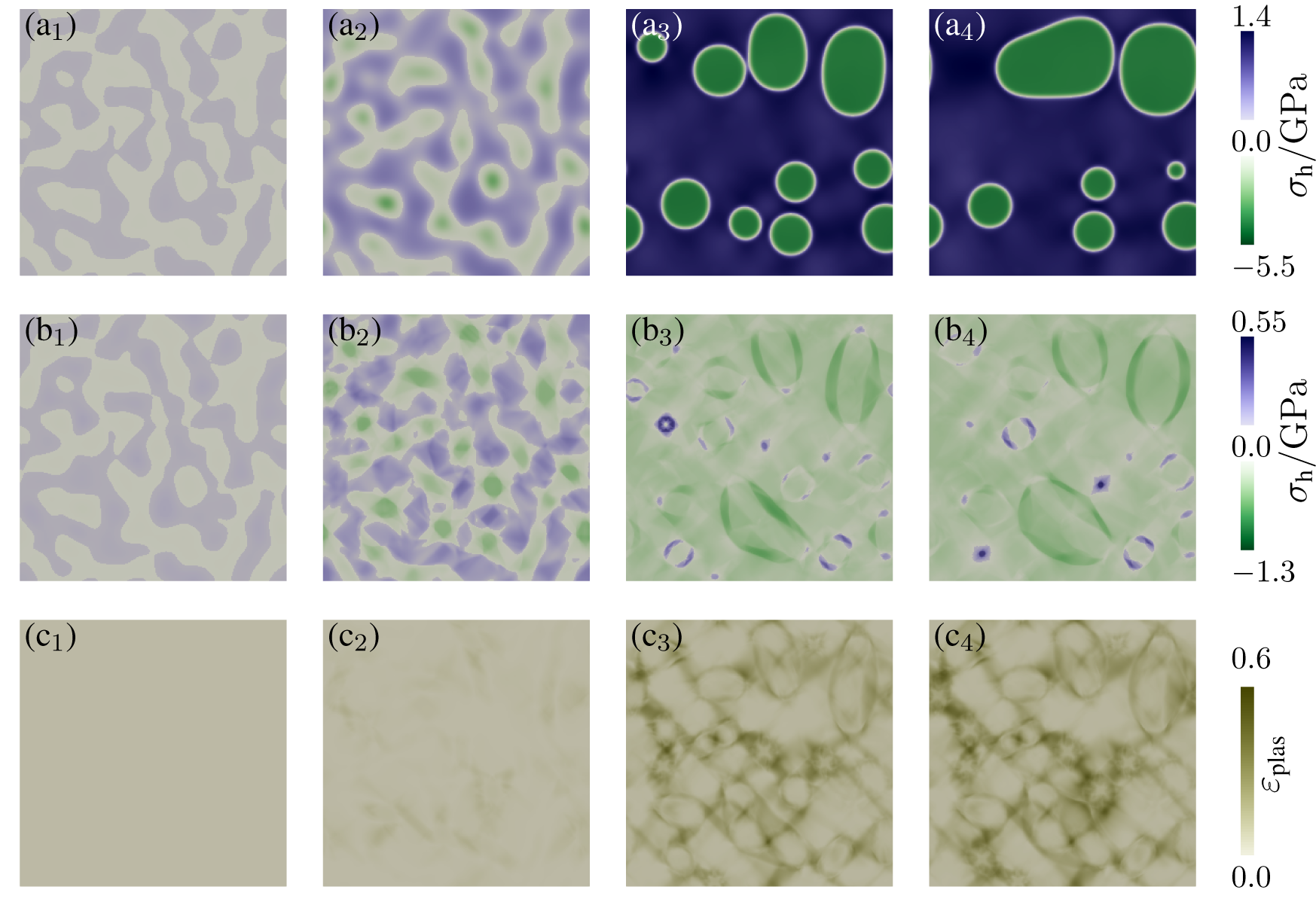}\end{overpic}
	\end{tabular}
	\caption{%
		Evolution of the hydrostatic stress during spinodal decomposition with (a) elastic, and (b) elasto-viscoplastic accommodation of the deformation due to the volumetric mismatch between solute components, and (c) the corresponding evolution of the plastic strain, for a mismatch strain coefficients of $\nu_{1} =0.03$, at time (1) 500s, (2) 800s, (3) 6000s, and (4) 10000s.
	}
	\label{fig: 2D_chem_stress_shearplas}
\end{figure}

\Cref{fig: 2D_chem_stress_shearplas} presents the evolution of the hydrostatic stress and plastic strain for the cases with elastic-only and elasto-plastic deformation at a volumetric mismatch coefficient of 0.03.
The plastic deformation relaxes the hydrostatic stress generated due to solute agglomeration during spinodal decomposition, and thus dissipates a significant portion of the stored elastic energy.
The plastic strain localization surrounding the solute-rich regions is observed, with a maximum value of 0.6.
More interestingly, even though the spinodal compositions are similar to the case without mechanical deformation, the morphology of the solute-rich region differs significantly.
When considering elasto-plastic deformation, the solute-rich regions exhibit an elliptic morphology at the intermediate level of volumetric misfit, as shown in \cref{fig: 2D_chem_plas} ($\text{b}_\text{4}$).
At the high levels of volumetric misfit, the solute-rich regions merge together and multiple lamellar regions can be observed, as shown in \cref{fig: 2D_chem_plas} ($\text{c}_\text{4}$).

\subsection{Ternary chemo-mechanical spinodal decomposition}
\label{sec: three-dimensional}
In order to illustrate the applicability of the modelling approach to more complex material systems,
a large three-dimensional (3D) simulation of a chemo-mechanically coupled ternary spinodal decomposition and coarsening process is considered next.
The cubic domain, $\domain B_0 = [0,L] \times [0,L] \times [0,L]$, of size $L = \SI{1.28}{\micro \meter}$ is meshed with $128\times128\times128$ regular hexahedral elements.
The initial  concentrations for solute A and B are 0.3 and 0.3, respectively, with a random perturbation of amplitude 0.01.
The chemical free energy parameters are listed in \cref{Tab: matepara3D}, and the crystal plasticity material parameters are listed in \cref{Tab: matepara2D}. 
The chemical free energy parameters used will result in the formation of three decomposition regimes: (A-rich, B-poor), (A-poor, B-rich), and (A-poor, B-poor).

\begin{table}[h!]
	\centering
	\caption{Chemical free energy parameters for the ternary spinodal system. $\Omega$ is the molar volume; $E_A^{sol}$ and $E_B^{sol}$ are the solution energies, $\kappa_A$ and $\kappa_B$ are the gradient coefficients, $\alpha_A$ and $\alpha_B$ are the penalty parameters, and ${M}_{A}$ and ${M}_{B}$ are the mobilities for component $A$ and $B$ respectively;  $E_{AA}^{int}$  and  $E_{AB}^{int}$ are the $A-A$ and $A-B$ interaction energies; and $\theta$ is the temperature.}
	\resizebox{0.9\linewidth}{!}{%
		\begin{tabular}{ccccccccccc}
			\toprule
			\text{Parameter} & \text{Value} &\text{Parameter} & \text{Value}\\
			\midrule		
			 $\Omega$ & $1\times10^{-5}$ ($\text{m}^3 \text{mol}^{-1}$)        & $\kappa_A$ & $1\times10^{-18}$ ($\text{J}\text{m}^{2}\text{mol}^{-1}$) \\
			 $E_A^{sol}$ & $1.24\times10^{4}$ ($\text{J} \text{mol}^{-1}$)     & $\kappa_B$ & $1\times10^{-18}$ ($\text{J}\text{m}^{2}\text{mol}^{-1}$) \\
			 $E_B^{sol}$ & $1.24\times10^{4}$ ($\text{J} \text{mol}^{-1}$)     & $\alpha_A$ & $1\times10^{5}$ ($\text{J} \text{mol}^{-1}$)\\
			 $E_{AA}^{int}$ & $-1.24\times10^{4}$ ($\text{J} \text{mol}^{-1}$) & $\alpha_B$ & $1\times10^{5}$ ($\text{J} \text{mol}^{-1}$)\\
		     $E_{AB}^{int}$ & $-1.24\times10^{4}$ ($\text{J} \text{mol}^{-1}$) & $M_{A}$ & $2.2\times10^{-19}$ ($\text{m}^{5} \text{s}^{-1} \text{J}^{-1}$) \\
			 $\theta$ & 498(K)                                                 & ${M}_{B}$ & $2.2\times10^{-19}$ ($\text{m}^{5} \text{s}^{-1} \text{J}^{-1}$)\\    
			\bottomrule
	\end{tabular}}
	\label{Tab: matepara3D}
\end{table}

\begin{figure}
	\centering
	\begin{tabular}{lcrl}
		\begin{overpic}[width=0.9\textwidth]{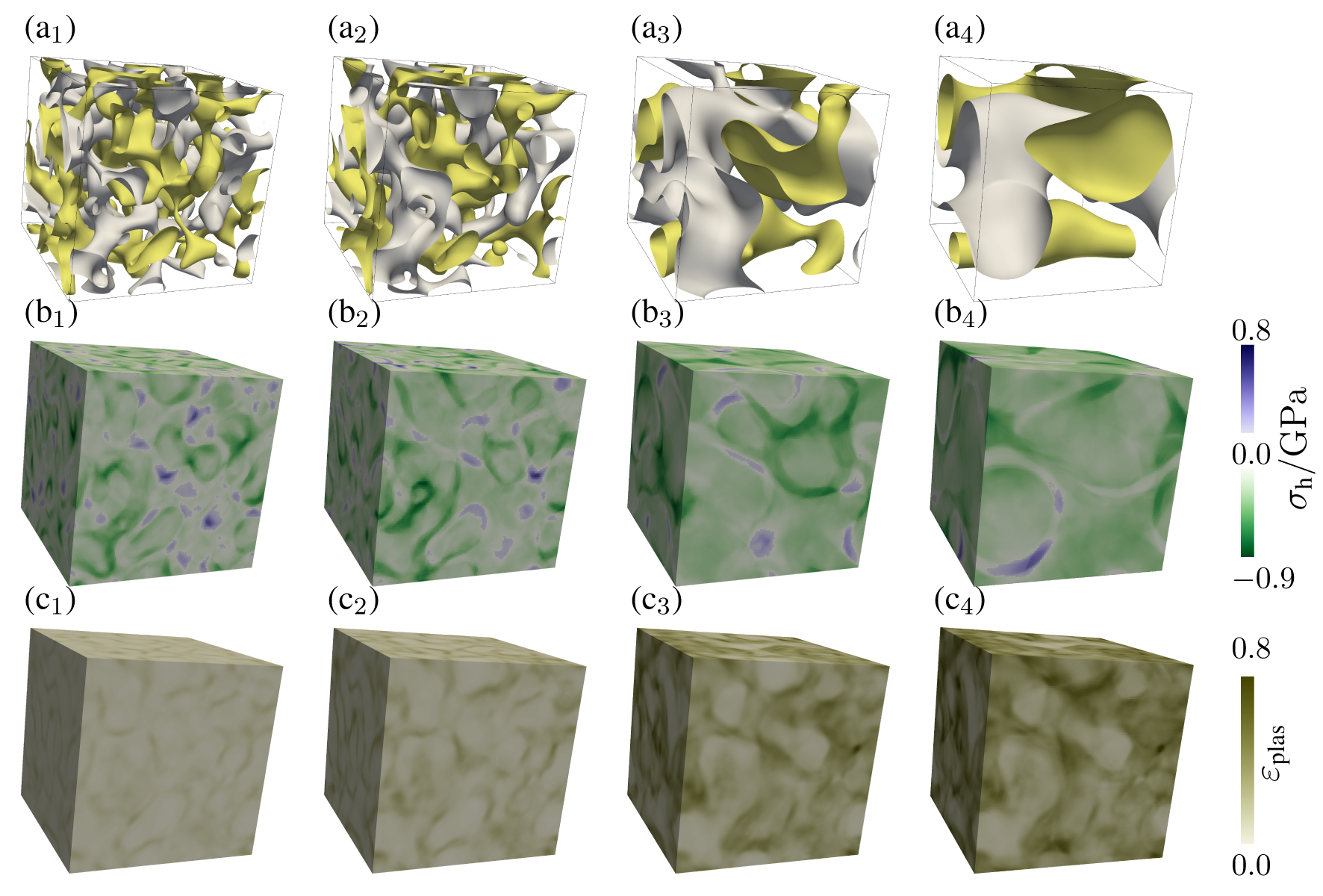}\end{overpic}
	\end{tabular}
	\caption{%
		Evolution of (a) the morphologies of the decomposition regions, (b) the hydrostatic stress, and (c) the plastic strain during a ternary spinodal decomposition and coarsening process, in a crystalline anisotropic elasto-plastic deforming material, with a mismatch strain coefficient of $\nu_{1} =0.03$, at time (1) 400s, (2) 1000s, (3) 6000s, and (4) 15000s. The grey and yellow iso-surfaces represent (A-rich, B-poor) and (A-poor, B-rich) regions respectively, within the (A-poor, B-poor) matrix. The concentrations of the iso-surfaces for both solute A and B are taken as 0.9.
	}
	\label{fig: 3D_plastic}
\end{figure}

\Cref{fig: 3D_plastic} shows the temporal evolution of the morphologies of different decomposition regimes, the accompanying hydrostatic stress, and the plastic strain.
In this fully coupled 3D ternary spinodal-elastic-viscoplastic mechanical decomposition simulation, the separated regions exhibit inter-connected morphologies instead of the isolated islands like those in the 2D binary cases,
as the three components have relatively the same concentrations (0.3, 0.3, 0.4), close to the off-critical point.
The hydrostatic stress distribution is significantly heterogeneous, even inside the same decomposition region, due to the accompanying plastic deformation, as shown in \cref{fig: 3D_plastic}(b).
It can be seen that the magnitude of the hydrostatic stress only changes slightly during the phase coarsening stage, while the total plastic strain increases gradually (\cref{fig: 3D_plastic}(c)).

\begin{figure}
	\centering
	\begin{tabular}{lcrl}
		\begin{overpic}[width=0.9\textwidth]{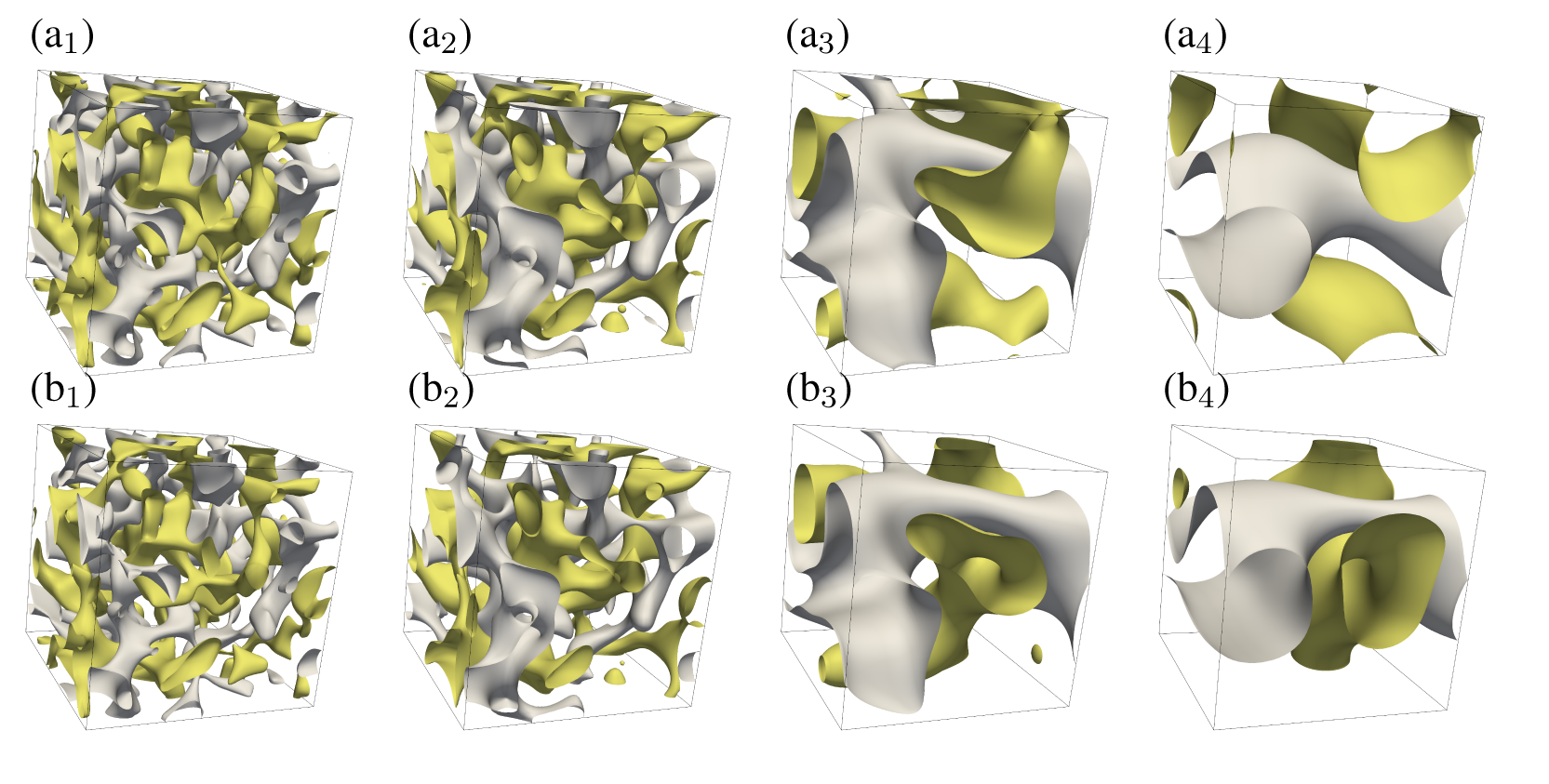}\end{overpic}
	\end{tabular}
	\caption{%
		Evolution of the spinodal decomposition and coarsening in a ternary spinodal decomposition and coarsening process, (a) without volumetric mismatch between solute components, and (b) considering elastic accommodation of the volumetric mismatch between solute components, with a volumetric mismatch coefficient of $\nu_{1} =0.03$, at time (1) 400s, (2) 1000s, (3) 6000s, and (4) 15000s. The grey and yellow iso-surfaces represent (A-rich, B-poor) and (A-poor, B-rich) decomposition regions respectively, and the rest is (A-poor, B-poor) matrix. The concentrations of the iso-surfaces for both solute A and B are taken as 0.9.
	}
	\label{fig: 3D_chem_elas_plas}
\end{figure}

The corresponding 3D simulations of the ternary spinodal decomposition in the absence of mechanical deformation and only considering elastic deformation are presented in \cref{fig: 3D_chem_elas_plas}.
Comparing \cref{fig: 3D_plastic}(a) and \cref{fig: 3D_chem_elas_plas}, the separated regions produced by these three simulations exhibit qualitatively similar inter-connected morphologies.
The approach of \citep{Gameiro2005} could be used to quantify and distinguish the geometrical differences between these complicated patterns, 
however, a full discussion is beyond the scope of the current work and represents work in progress to be reported in the future.

\section{Conclusions and Outlook}
In this work, a numerical approach has been introduced, using a semi-implicit convex splitting to invert the multi-component chemical potential relations, and resulting in a numerically advantageous expression for their transport in terms of the chemical potential. 
The approach is validated using a generalized spinodal Landau-type system as a benchmark, exposed to different chemo-mechanical starting and boundary conditions.
A convergence of the proposed weakly non-local approach to the strong CH form is demonstrated with increasing values of the penalty parameter.
However, for practical purposes, a penalty parameter of $1\times 10^5 \text{J} \text{mol}^{-1}$ was found to give sufficiently converged results.
A significant reduction of the numerical cost and stability of the proposed approach is also demonstrated, allowing for the use of larger time steps in long-term diffusion simulations.

The influence of chemo-mechanical coupling, including both elastic and anisotropic crystalline plastic deformation, on the spinodal decomposition and coarsening process is also investigated.
It is found that the stored elastic energy can narrow the miscibility gap and alter the spinodal compositions.
The decomposition kinetics is significantly reduced with increasing volumetric mismatch between solute components
Furthermore, the morphology of the solute-rich phase changes from spherical shape distributions to an ordered cubic distribution due to the anisotropy of the elastic stiffness tensor. 
When taking in to account additional visco-plastic relaxation of the volumetric mismatch, as in the case of realistic materials applications, it is found that the plastic relaxation accounts for a significant amount of the dissipated total energy.
The spinodal compositions are close to those in the absence of mechanical deformation, but the solute-rich regions tend to coalesce into elongated lamellar morphologies.

Extension of the proposed approach to multi-phase systems is straightforward.
This will be interesting as, in the context of multiphase systems, working directly with the chemical potential lends itself naturally to the Kim-Kim-Suzuki (KKS) description of interfaces \citep{Kim1999}, since, algorithmically, no partitioning of an average solute composition into the individual phases is required.
The proposed approach is also particularly amenable to CALPHAD-based descriptions of the chemical energy \citep{Lukas2007}, which is essential in the drive towards predictive simulations.   

\section{Acknowledgements}
PS CL and BS gratefully acknowledge the financial support by the DFG through the Priority Programme SPP 1713: Strong Coupling of Thermo-chemical and Thermo-mechanical States in Applied Materials.
PS is also grateful to  the EPSRC for financial support through the associated programme grant LightFORM (EP/R001715/1) and the Airbus\textendash University of Manchester Centre for Metallurgical Excellence for supporting aspects of this research.

\section{References}

\bibliographystyle{elsarticle-harv}

\bibliography{Shanthraj_etal2018}

\clearpage
\end{document}